\documentclass[]{aa}

\usepackage{graphicx}
\usepackage{amsmath,amssymb}
 											
\newcommand{\Teff}{${T_\mathrm{eff}}$}
\newcommand{\FeH}{$\mathrm{[Fe/H]}$}

\newcommand{\logg}{$\log~\mathrm{g}$}
\newcommand{\g}{$\mathrm{g}$}
\newcommand{\Y}{$Y$}
\newcommand{\X}{$X$}
\newcommand{\Z}{$Z$}
\newcommand{\BV}{$B-V$}
\newcommand{\Mbol}{${M_{\mathrm{bol}}}$}

\newcommand{\aMLT}{${\alpha_\mathrm{MLT}}$}
\newcommand{\aMLTs}{$\alpha_{\mathrm{MLT,} \odot}$}

\newcommand{\Msun}{$\mathrm{M_{\odot}}$}

\newcommand{\Mv}{$M_{\mathrm{V}}$}
\newcommand{\dydz}{{$\Delta Y/\Delta Z$}~}
\newcommand{\dydzsun}{${(\Delta Y/\Delta Z)}_\odot$}
\newcommand{\Yp}{${Y_\mathrm{p}}$}
\newcommand{\etal}{et~al.\ }
\newcommand{\V}{$V$}
\newcommand{\Hp}{$Hp$}

\begin{document}


\title{The Helium content and age of the Hyades:}

\subtitle{Constraints from five binary systems and Hipparcos parallaxes}

\author{Yveline Lebreton\inst{1} \and Jo\~ao Fernandes \inst{2}
\and Thibault Lejeune \inst{2}}

\offprints{Y. Lebreton,
\email{Yveline.Lebreton@obspm.fr}}

\institute{DASGAL, CNRS UMR 8633, Observatoire de Paris, Place J.  Janssen, 92195
Meudon,  France  \and Observat\'orio Astron\'omico da
Universidade de Coimbra, 3040 Coimbra, Portugal}

\titlerunning{The helium content and age of the Hyades}
\authorrunning{Y. Lebreton et al.}

\date{Received 4 April 2001/Accepted 17 May 2001}

\abstract{
We compare the accurate empirical mass-luminosity (M-L) relation based on five Hyades binary 
systems to predictions of stellar models calculated with various input parameters
(helium, metallicity and age) or physics (mixing-length ratio, model 
atmosphere, equation of state and microscopic diffusion). Models based on a helium content
\Y$\sim$0.28 inferred from the \dydz enrichment law are more than 3$\sigma$ 
beyond the observations, suggesting that
the Hyades initial helium abundance is lower than expected from its supersolar metallicity.
With the photometric metallicity (\FeH=0.144$\pm$0.013 dex, Grenon \cite{gre00})
we derive \Y=0.255$\pm$0.009. Because of the (\Y, \FeH) degeneracy in the M-L plane,
the uncertainty grows to $\Delta Y=0.013$ if the metallicity from spectroscopy is adopted
(\FeH=0.14$\pm$0.05 dex, Cayrel de Strobel \etal \cite{cayG97}). 
We use these results to discuss the Hertzsprung-Russell (HR) diagram of the 
Hyades, in the (\Mv, \BV) plane, based on the very precise Hipparcos dynamical parallaxes. 
Present models fit the tight observed sequence very well except at low temperatures. We show 
that the HR diagram does not bring further constraints on the helium abundance
or metallicity of the cluster. In the low mass region of the HR diagram sensitive 
to the mixing-length parameter (\aMLT), the slope of the main sequence (MS)
suggests that \aMLT\ could decrease from a solar (or even supersolar)
value at higher mass to subsolar values at low mass, which is also supported by the modeling of 
the vB22 M-L relation. We find that the discrepancy at 
low temperatures ((\BV) $\gtrsim$ 1.2) remains, even if an improved equation of state
or better model atmospheres are used. Finally, we discuss the positions of
the stars at turn-off 
in the light of their observed rotation rates and we deduce that the maximum age of the Hyades predicted by
the present models is $\sim$650 Myr.
\keywords{open clusters and associations: individual: Hyades -- stars:  fundamental parameters -- stars: interiors
-- stars:  Hertzsprung-Russell (HR) and C-M diagram -- stars: individual:
vB22, $\theta^2$~Tau - stars: rotation}
}

\maketitle

\section{Introduction}

Open clusters provide information and strong constraints for the
stellar evolution theory. They give the opportunity to study large numbers 
of stars spanning a broad range of masses and evolutionary stages and that
can be assumed to have similar age and chemical composition.
Depending on the cluster studied, observations may give 1) the position of the cluster sequence 
in the HR diagram, 2) the density of stars along that sequence, 
3) the  M-L relation if the masses of some binary stars
are accessible to observation. The analysis of the observational 
features of a given cluster by means of internal structure models allows estimating 
characteristics not directly accessible through observation, such as the age or helium 
content of the members. Furthermore, if the observational data are accurate enough, 
constraints on the physical processes at work in the stellar interiors, for
instance the various transport processes, can be inferred.					  

The Hyades is the nearest moderately rich star cluster. It has served for a
long time to define absolute magnitude calibrations and, in turn,
to fix the zero-point of the galactic and extragalactic distance scales.
Also, as an open cluster, the precise knowledge of its chemical composition
and age is fundamental for studies of the kinematic
and chemical evolution of our Galaxy. The metallicity \FeH~ (logarithm of the number abundances of 
iron to hydrogen relative to the solar value) is accessible
through photometric or spectroscopic observations (see Sect.~\ref{obs}). On the other
hand, the Hyades dwarfs are too cool for helium lines to
be visible in their spectra and their helium content to be determined directly.
The helium abundance and  age have to be derived from the analysis 
of the observations using stellar models.

Recently, high-quality observations of the Hyades 
stars have been obtained. Hipparcos data combined with ground-based 
photometric or spectroscopic observations provided a more
precise extended HR diagram of the cluster
(Perryman \etal \cite{per98}, Dravins \etal \cite{dra97}, de
Bruijne \etal \cite{bru00}).
On the other hand, observations of several binary systems in the Hyades
 yielded 
a much improved M-L relation
(Torres \etal \cite{tor97a}, \cite{tor97b}, 
\cite{tor97c}; Peterson \& Solensky \cite{pet88};
Söderhjelm \cite{sod99}). 

Perryman \etal (\cite{per98}), Lebreton \etal (\cite{leb97}) and 
Cayrel de Strobel \etal (\cite{cayG97}) analysed the observational 
HR diagram of the Hyades with stellar models and obtained
new estimates of the age, initial helium content and metallicity
of the cluster (see Sect.~\ref{obs}). Recently, de Bruijne \etal (\cite{bru00})
and Castellani \etal (\cite{cas00}) compared the observational HR diagram with models
and discussed the different uncertainties, 
in particular the problems related to the color-magnitude calibrations.

In this paper, we focus on the determination of the helium abundance
of the Hyades through the analysis of both the M-L relation and HR diagram. We show that
the excellent accuracies on the masses and luminosities	reached for 
the Hyades binaries provide rather severe constraints on the helium
content of the cluster and allow confirming and refining of the 
helium value previously inferred from the analysis of the HR diagram. In a
second step, we discuss our ability to reproduce the various features of the HR diagram 
cluster sequence with different model input physics.

Sect.~\ref{H-R} is a brief review of previous  
studies of the observed HR diagram of the Hyades. 
Sect.~\ref{M-L} presents the observed Hyades M-L relation.
Numerical stellar models are presented in Sect.~\ref{models}.
In Sect.~\ref{helium}, we analyse the M-L relation
by means of stellar models and discuss the implications for
the chemical composition of the cluster. In Sect.~\ref{HR},
we analyse and discuss the HR diagram of the cluster on the basis of 
the constraints provided by the M-L relation. Summary and conclusions are given 
in Sect.~\ref{summary}.

\begin{figure}[htb]
\resizebox{\hsize}{!}{\includegraphics*{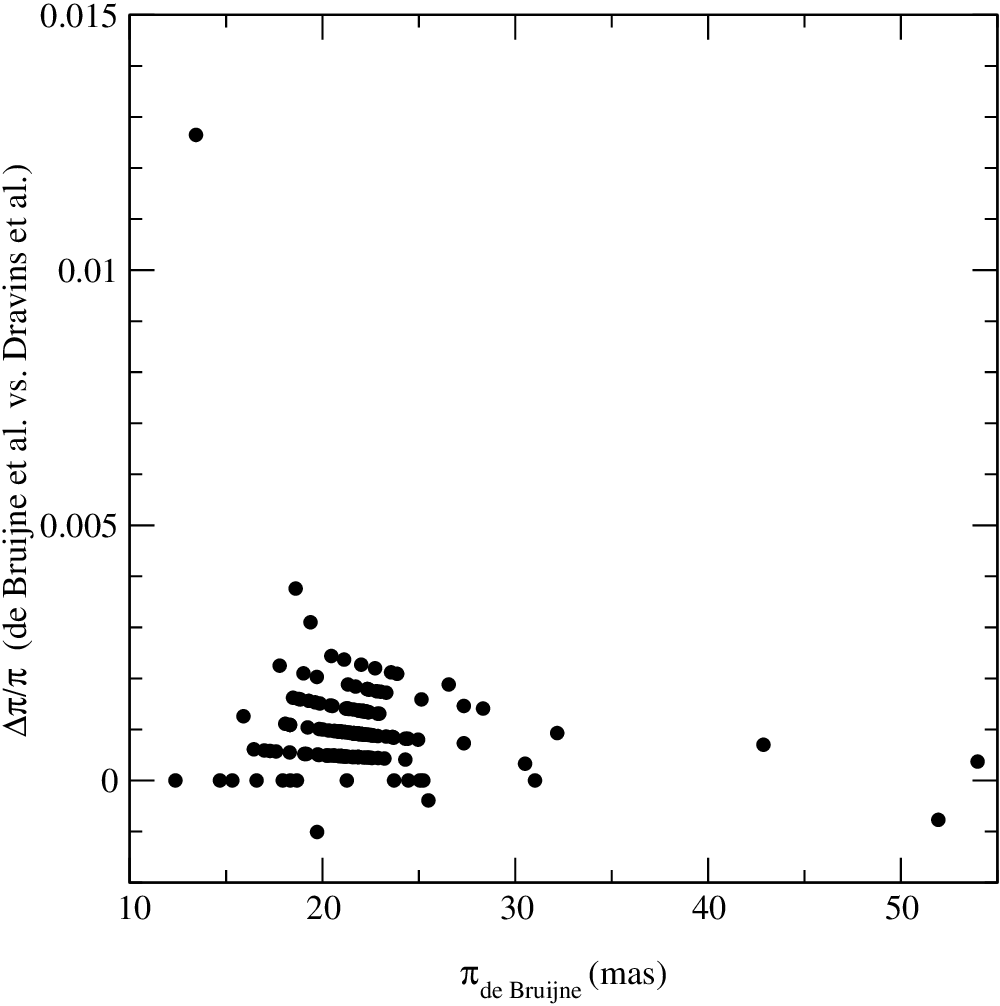}}
\caption{Comparison of dynamical parallaxes of Hyades' stars
as determined by de Bruijne \etal 
(\cite{bru00}) and by Dravins
\etal (\cite{dra97}): all stars but one (HIP 28356) have their parallaxes
within less than 0.4 percent}
\label{pi}
\end{figure}
\section{HR diagram data and previous analyses} 
\label{H-R}

\subsection{Observational data} 
\label{obs}

Hipparcos measured the positions, proper motions
and trigonometric parallaxes of $\sim$300 candidate Hyades members
allowing new, more detailed studies of the cluster to be undertaken. 

Perryman \etal (\cite{per98}) assigned the membership 
of 218 cluster stars and from the individual trigonometric parallaxes 
of the 134 stars located within 10 pc of the cluster center, 
obtained a distance modulus $(m-M) = 3.33 \pm 0.01\ \mathrm{mag}$.
Narayanan \& Gould (\cite{nar99}) used the Hipparcos proper 
motions to derive statistical parallaxes for 43 cluster stars leading to
$(m-M) = 3.34 \pm 0.02\ \mathrm{mag}$. 

Dravins \etal (\cite{dra97}), and more recently
de Bruijne \etal (\cite{bru00}), derived the dynamical parallaxes of $\sim$200
Hyades members from the relation between the cluster space motion,
the positions and the projected proper motions. As shown in Fig.~\ref{pi}, the two groups, 
who used the same data and similar methods, obtain parallaxes in excellent agreement; the dynamical
parallaxes of the stars they have in common agree to better than 0.4 percent (except for one star, 
HIP 28356, where the parallax difference is about 1 percent). The mean accuracy on the
dynamical parallaxes is $\sim$0.5 mas, that is $\sim$3 times better than the
trigonometric parallax accuracy.

Photometric information can be found in the Hipparcos catalogue. The Hipparcos 
\Hp-magnitude of each observed star is given with an accuracy of 0.0015
mag. The \V-magnitude and (\BV) index of each star coming from ground-based measurements
are also given in the catalogue, the typical accuracy on \V\  is better than 0.01 mag.

The mean absolute magnitude accuracy of the Hyades
stars based on the dynamical parallaxes of Dravins \etal
(\cite{dra97}) or of de Bruijne \etal (\cite{bru00}) is $\sigma_{M_V} \simeq 0.05$ mag.
De Bruijne \etal (\cite{bru00}) examined 218 Hyades candidates
among which they selected 90 secure Hyades members.  The resulting (\Mv, \BV) colour-absolute magnitude
diagram shows a very well-defined MS (i.e. $\langle \sigma_{M_V}
\rangle \simeq 0.05$ mag and $\langle \sigma_{(B-V)} \rangle \simeq
0.01$ mag, see Fig.~\ref{diag_HR}).

The analysis of high-resolution, high signal-to-noise spectra of several Hyades stars
has provided precise determinations of their effective temperatures, \Teff, 
and metallicities, \FeH. Cayrel de Strobel \etal
(\cite{cayG97}) selected 40 Hyades dwarfs with \Teff~ accurate to
typically 50-70 K; their mean metallicity \FeH = $+0.14 \pm 0.05$ dex, is in good agreement with the photometric
value \FeH = $+0.144 \pm 0.013$ dex derived by  Grenon (\cite{gre00})
from large sets of homogeneous observations
in the Geneva photometric system. 

\subsection{Previous analyses by means of stellar models} 

Cayrel de Strobel \etal (\cite{cayG97}) obtained the lower part of the cluster MS, in the (\Mbol, \Teff) plane,
from the positions of 40 dwarfs, by combining Hipparcos distances, 
spectroscopic \Teff, \V-magnitudes from the Hipparcos 
input catalogue and bolometric corrections by Bessel \etal (\cite{bes98}).
Lebreton \etal (\cite{leb97}) and Perryman \etal (\cite{per98})
compared the bottom of this sequence, corresponding to the
non-evolved stars, with theoretical zero-age main sequences (ZAMS)
 computed with a $Z/X$ ratio of $0.034$ corresponding to
the mean observed \FeH-value \footnote{\Z~and \X~are respectively the abundances in mass of
heavy elements and hydrogen related to \FeH~through 
\FeH=$\log (Z/X) -\log (Z/X)_\odot$ where
$(Z/X)_\odot=0.0245$ according to Grevesse \& Noels (\cite{gre93}).}. 
The ZAMS fitting yielded the initial 					
helium content \Y= 0.26$\pm$0.02 (in mass fraction) and metallicity \Z= 0.024$\pm$0.004
of the cluster, the (quite large) error on $Y$ being dominated by
the error on the mean value of \FeH~used (\FeH=$+0.14 \pm 0.05\ \mathrm{dex}$ from spectroscopy).

The position of the whole cluster sequence in the (\Mv, \BV) plane
was obtained by Perryman \etal (\cite{per98}) from the combination
of Hipparcos distances and ground-based \V-magnitudes and 
(\BV) color indexes. The fitting of the sequence by model
isochrones corresponding to the chemical composition of the cluster 
(\Y= 0.26, \Z= 0.024) yielded an estimate of the age, $t \simeq 625$ Myr.

In order to obtain a MS as fine as possible, the known or suspected
unresolved binaries were removed from the samples mentioned above
and the masses of the few resolved binaries were not used 
as constraints. Binaries with well-determined masses provide valuable constraints
for stellar models, in particular for the stellar abundances through the M-L relation 
(see e.g. Andersen \cite{and91}). Some difficulties in fitting parts of the
empirical Hyades M-L relation with models have been encountered 
by Torres \etal (\cite{tor97b}) and Lastennet \etal (\cite{las99}), but the models (available 
from the literature) did not correspond to the cluster chemical composition.
We now turn to the sample of binaries well-observed in the Hyades.

\begin{table*}
\caption[]{Parallaxes and masses for the 5 binary systems. The dynamical parallaxes $\pi_\mathrm{
dyn,1}$ are from Lindegren (1999, private communication but see also Dravins
et al. \cite{dra97}). The dynamical parallaxes $\pi_\mathrm{
dyn,2}$ are from de Bruijne \etal  (\cite{bru00}). The Hipparcos trigonometrical parallaxes
$\pi_\mathrm{trig}$ come from Perryman \etal (\cite{per98}). As discussed in the text, 
the orbital parallaxes $\pi_\mathrm{orb}$ and individual masses were obtained by Peterson \& Solensky 
(\cite{pet88}) for vB22 and by Torres \etal 
(\cite{tor97a}, \cite{tor97b}, \cite{tor97c}) for the other systems. 
The radii of the vB22 components are from Schiller \& Milone (\cite{sch87}) where the probable error has been
translated into a mean error.}

\begin{tabular}{ccccccccc}
\hline\noalign{\smallskip}
name& HIP   & HD &$\pi_\mathrm{dyn, 1}$  &$\pi_\mathrm{dyn, 2}$&$\pi_\mathrm{trig}$
& $\pi_\mathrm{orb}$&$\mathrm{M_A/M_\odot}$ &$\mathrm{R_A/R_\odot}$  \\
          &          &          &  (mas)    &  (mas)     &  (mas) & (mas)  &$\mathrm{M_B/M_\odot}$& 
$\mathrm{R_B/R_\odot}$   \\
\noalign{\smallskip}
\hline\noalign{\smallskip}
vB22            &20019 & 27130 & 21.15$\pm$0.33 & 21.16$\pm$0.38 &21.40$\pm$1.24 & $19.8\pm0.4$ &$1.072\pm0.010$&$0.905 \pm0.029$\\ 
                &      &       &&                &               &        &$0.769\pm0.005$& $0.773 \pm0.015$\\
51 Tau          &20087 & 27176 & 18.29$\pm$0.24 & 18.31$\pm$0.69 & 18.25$\pm$0.82& 17.9 $\pm$0.6 &$1.80 \pm0.13$ & \\
                &      &       &      &          &               &               &$1.46 \pm 0.18$& \\
$\phi$~342      &20661 & 27991 & 21.27$\pm$0.33 & 21.29$\pm$0.37 & 21.47$\pm$0.97& 21.44$\pm$0.67&$1.363 \pm0.073$&\\
              &      &       &      &          &               &               &$1.253 \pm0.075$ & \\
$\theta^{1}$ Tau&20885 & 28307 &        -       & 21.29$\pm$0.37 & 20.66$\pm$0.85& -             &$2.77  \pm0.50$&\\ 
                &      &       &       &         &               &               &$1.28  \pm0.13$&\\
$\theta^{2}$ Tau&20894 & 28319 & 22.21$\pm$0.35 & 22.24$\pm$0.36 & 21.89$\pm$0.83& 21.22$\pm$0.76&$2.42\pm0.30$&\\ 
                &      &       &           &     &               &               &$2.11\pm0.17$&\\
\noalign{\smallskip}
\hline
\end{tabular}
\label{systems}
\end{table*}
\begin{table*}
\caption[]{Magnitudes and colour indices for the five binary systems.}

\begin{tabular}{ccccccc}
\hline\noalign{\smallskip}
name& HIP   & HD   &$V$       &$M_\mathrm{V,A}$ &\BV&$(B-V)_\mathrm{A}$\\
          &          &          &$\Delta V$&$M_\mathrm{V,B}$ &$\Delta$(\BV)&$(B-V)_\mathrm{B}$\\
\noalign{\smallskip}
\hline\noalign{\smallskip}
vB22            &20019 & 27130 &$8.319\pm0.009$&$5.07\pm0.04$&     -       &$0.713\pm0.017$\\ 
                &      &       &$2.3\pm0.05   $&$7.37\pm0.06$&     -       &$1.19\pm0.08$\\
51 Tau          &20087 & 27176 &$5.65\pm0.01$  &$2.18\pm0.09$&$0.28\pm0.01$&$0.25\pm0.05$\\
                &      &       &$1.61\pm0.10$  &$3.79\pm0.12$&$0.19\pm0.01$&$0.44\pm0.02$ \\
$\phi$~342      &20661 & 27991 &$6.46\pm0.01$  &$3.70\pm0.05$&$0.49\pm0.01$&$0.466\pm0.056$\\ 
                &      &       &$0.34\pm0.05$  &$4.03\pm0.05$&$0.05\pm0.01$&$0.530\pm0.020$\\
$\theta^{1}$ Tau&20885 & 28307 &$3.84\pm0.01$  &$0.52\pm0.04$&$0.95\pm0.01$&-\\ 
                &      &       &$3.50\pm0.05$  &$4.02\pm0.07$&   -   &     -  \\
$\theta^{2}$ Tau&20894 & 28319 &$3.40\pm0.02$  &$0.48\pm0.04$&$0.18\pm0.01$&$0.18\pm0.02$\\ 
                &      &       &$1.10\pm0.01$  &$1.58\pm0.05$&$-0.006\pm0.005$&$0.17\pm0.02$\\
\noalign{\smallskip}
\hline
\end{tabular}

\label{magnitudes}
\end{table*}

\section{Updated data for five binary systems}
\label{M-L}

We focus on five Hyades binary systems with well-determined masses.
The data are listed in Table~\ref{systems}. The dynamical parallaxes, $\pi_\mathrm{dyn}$, 
are more precise than the direct Hipparcos trigonometrical parallaxes, $\pi_\mathrm{trig}$ 
(Perryman \etal \cite{per98}). Therefore, in the following, we have used the dynamical parallaxes of de Bruijne \etal
(\cite{bru00}) available for the five systems. We also list in Table~\ref{systems} the $\pi_\mathrm{dyn}$-values
derived by Dravins \etal (\cite{dra97})
in their improved version (Lindegren 1999, private communication).
  
For three systems (51 Tauri, Finsen 342 and $\theta^{2}$ Tau),
which are double-lined spectroscopic binaries as well as visual binaries
resolved by speckle, Torres \etal (\cite{tor97a}, \cite{tor97b}, 
\cite{tor97c}) derived a complete astrometric-spectroscopic orbital solution, and
therefore could obtain the individual masses and the orbital parallax, $\pi_\mathrm{orb}$, 
listed in Table~\ref{systems}. 51 Tauri is also one of the 25 binary systems for which 
Söderhjelm (\cite{sod99}) could derive the individual masses combining Hipparcos data with
ground-based observations. The resulting masses,
$\mathrm{M_A}=1.72\pm0.27\ \mathrm{M_\odot}$ for the primary and $\mathrm{M_B=1.31\pm0.21\ M_\odot}$ for the 
secondary are in good agreement with those obtained by Torres et al. Also, Martin \etal (\cite{mar98}) 
obtained the masses of the two components of 51 Tau from the analysis of Hipparcos data but
with a lower accuracy ($\mathrm{M_A=1.756\pm0.343\ M_\odot}$ and $\mathrm{M_B=0.953\pm0.247\ M_\odot}$). 

For the $\theta^1$ Tau system,  a single-lined spectroscopic binary with astrometric 
information from speckle and lunar occultation measurements, Torres \etal (\cite{tor97c}) 
obtained a partial astrometric-spectroscopic solution where the information coming from
the velocity amplitude of the B-component was lacking. However, it is possible to estimate
the individual masses, using the parallax obtained independently from Hipparcos data.
We give in Table~\ref{systems} the values of the masses of the components of $\theta^1$ 
derived from formulae (2), in Torres et al (\cite{tor97c}), using their orbital parameters 
together with the Hipparcos $\pi_\mathrm{dyn, 2}$ dynamical parallax. It is worth 
noting that $\theta^1$ Tau B and $\phi$342B have very similar masses and magnitudes.

The fifth system, vB22, is a double-lined spectroscopic eclipsing binary discovered by McClure 
(\cite{mac82}). Individual masses were derived by Peterson \& Solensky (\cite{pet88}) and the radii 
were determined by Schiller \& Milone (\cite{sch87}).

The \V-magnitude of each system, the magnitude difference between the two components	
$\Delta V$ and the corresponding individual absolute magnitudes $M_\mathrm{V, A}$ and 
$M_\mathrm{V,B}$ (corresponding to the $\pi_\mathrm{dyn, 2}$-values) are listed in Table~\ref{magnitudes}.
The individual values of the (\BV) color index of the two components of vB22 have been determined by 
Schiller \& Milone (\cite{sch87}). For the other systems we infer the individual values of (\BV) from
the global value of the system given in WEBDA ({\tt http://obswww.unige.ch/webda/}, see Mermilliod \cite{mer98}) 
and from the difference $\Delta$(\BV). For $\theta^2$ Tau, $\Delta$(\BV) has been measured by 
Peterson \etal (\cite{pet93}). For 51 Tau and $\phi$342, we follow 
Torres \etal (\cite{tor97a}) who derived $\Delta(B-V)$ from the value of $\Delta V$ and of 
the local slope of the empirical (\V, \BV) relation of Schwan \etal
(\cite{sch91}) for the Hyades main sequence. For the $\theta^1$ Tau system,
which is composed of a giant and of a dwarf, we only know the
system (\BV)-value.

\section{Stellar models and related input physics}
\label{models}

We have calculated stellar models in the mass range
$\mathrm{M \in~[0.5, 3.0]~M_\odot}$, using the CESAM code (Morel \cite{mor97})
and the input physics described below. 

\begin{enumerate}
\item Equation of state (EOS): the CEFF EOS (Eggleton \etal \cite{egg73}, Christensen-Dalsgaard \cite{chr91}) 
has mostly been used; it includes the Debye-H\"uckel corrections for pressure. 
The OPAL EOS (Rogers \etal \cite{rog96}) that 
includes collective effects has been used at low mass for the purpose of comparison. 
\item Opacities: we used OPAL (Iglesias \& Rogers \cite{igl96}) 
complemented by Alexander \& Ferguson's (\cite{ale94}) data for 
$T\lesssim10^4\ K$, both being for Grevesse \& Noels's 
(\cite{gre93}) solar mixture.
\item Thermonuclear reaction rates: Caughlan \& Fowler (\cite{cau88}).
\item Standard mixing-length theory of convection (B\"ohm-Vitense \cite{boh58}). 
The calibration of the solar model in luminosity and radius with a particular set of input physics 
yields the solar mixing-length ratio \aMLTs= $l/\mathrm{H_p}$ ($\mathrm{H_p}$ 
is the pressure scale-height). The value of \aMLTs\ 
depends on the external boundary conditions adopted. We find \aMLTs=1.61  for the Eddington's grey model 
atmosphere and \aMLTs=1.79 for Kurucz's 
(\cite{kur91}) ATLAS9 models (see item 7). Models calculated with \aMLTs\ also give
a good fit of the MS slope of the nearby stars (Lebreton \etal \cite{leb99}) and
of the Hyades (Perryman \etal \cite{per98}). In addition, the calibration in luminosity and
effective temperature of several nearby visual binaries with determined masses yields 
values of \aMLT\ close to \aMLTs\ (Fernandes \etal 1998).
From numerical 2-D simulations of convection in a wide range of \Teff\ and \logg\ (\g\ is the surface gravity), 
Ludwig \etal (\cite{lud99}) obtained equivalent \aMLT-values in the range 1.1-1.8 for solar or 
subsolar \FeH - they did not investigate supersolar metallicites as found in the Hyades. 
For {\emph{low mass MS stars}} these simulations suggest possible variations of
\aMLT\ with \Teff~ and \logg~
in the range \aMLTs~$\pm 0.15$. We therefore adopted the solar mixing-length ratio
for standard models but we discuss the effects of a change of \aMLT\ as high as $\pm 0.40$.
\item Convective core overshooting. We take $\mathrm{d_{ov}=0.2 H_p}$ as 
overshooting distance, a value 
derived by Schaller \etal (\cite{sch92}) from the study of the empirical MS width of open clusters.
We also discuss the case with no overshooting. 
\item Microscopic diffusion transports helium and heavy elements toward the center of stars and 
may change their surface and core
compositions during evolution. It can be important in rather old metal-poor stars
(see e.g. Proffitt \& Vandenberg \cite{pro91}). We calculated models including microscopic diffusion and
checked that in the Hyades, which are rather young, metal-rich objects,
the effects of microscopic diffusion are small enough to be neglected in the present study. 
\item External boundary conditions. The $T(\tau)$-law is taken 
from the classical Eddington grey model 
atmosphere  except in a few models where we used, for comparison, a $T(\tau)$-law derived from Kurucz's 
(\cite{kur91}) complete ATLAS9 model atmospheres.
\end{enumerate}

To fit the Hyades empirical M-L relation and HR diagram, we calculated 
evolution models and isochrones with various values of the initial helium abundance $Y$
and metallicity $Z/X$. 
The $Z/X$-ratio can vary in the range $0.034\pm0.005$ given by the observed \FeH-value
(see Sect.~\ref{H-R}). The error bar on $Z/X$ (0.005) includes the 0.05 dex error
on \FeH~ and the 11 per cent error on $(Z/X)_\odot$ quoted by Anders \& Grevesse (\cite{and89}).
The helium mass fraction $Y$ is allowed to vary in the range 0.235-0.30, i.e. from the
primordial abundance (see Peimbert \etal \cite{pei00}) to the approximative upper
limit given by observations (see Nissen \cite{nis76}). 
In the following, we refer to solar-scaled helium when the helium abundance
is related to metallicity through the well-known relation 
$(Y-Y_p)/Z$=\dydz=\dydzsun where \Yp=0.235 is the primordial helium value and
\dydzsun is derived from the calibration in luminosity and radius of the solar
model. For a solar model that does not take into account 
the microscopic diffusion we obtain
$Y_\odot=0.2674$ and $Z_\odot=0.0175$ and in turn \dydzsun=1.9.

The conversion of the theoretical model outputs (\Mbol, \Teff) into the
observational plane (\Mv, \BV) has been performed with the
most recent version of the BaseL Stellar Library (BaSeL, version 2.2),  
available electronically  at {\tt http://www.astro.mat.uc.pt/BaSeL/}
(Lejeune, \cite{lej01}).
BaSeL provides color-calibrated theoretical flux distributions for
the largest possible  range of  fundamental
stellar parameters, \Teff~ (2000 to 50,000  K), \logg~($-1.0$
to $5.5$ dex), and \FeH~($-5.0$ to $+1.0$ dex).

The BaSeL flux distributions are calibrated on the stellar
$UBVRIJHKL$ colors, using these {\em  empirical}  photometric
calibrations for solar metallicity, and {\em semi-empirical} relations
constructed from the color differences predicted  by the stellar model
atmospheres for non-solar metallicities (details about the calibration
procedure are given  in Lejeune \etal \cite{lej97}, \cite{lej98}).  

We point out that although Hipparcos provided its own precise magnitudes
$Hp$ (see Sect.~\ref{obs}), it has not been possible to use them here because
there are still problems in computing the bolometric corrections corresponding
to the $Hp$ band (Cayrel \etal \cite{cay97}).

All the models and isochrones presented here can be sent on request.

\section{Constraints from the M-L relation}
\label{helium}

\begin{figure}[htb]
\resizebox{\hsize}{!}{\includegraphics*{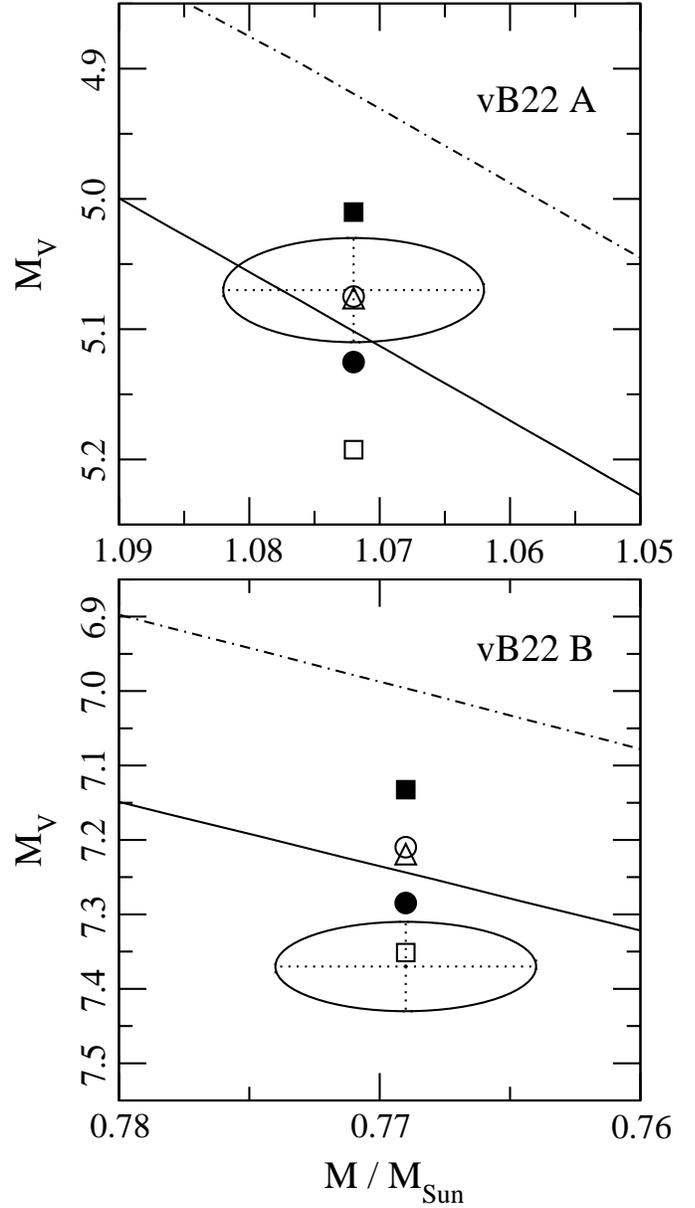}}
\caption{The vB22 system in the M-L plane (data from Tables~\ref{systems} \& \ref{magnitudes}).
The ellipses represent the locus where 
$(M_\mathrm{V}-M_\mathrm{V, obs})^2/{\sigma_{M_\mathrm{V}}}^2+(\mathrm{M-M_{obs}})^2/{\sigma_\mathrm{M}}^2 \leq 1$. 
Models aged 625 Myr are plotted.
Continuous line: three indistinguishable isochrones with (\FeH, \Y) $\equiv$ (0.09, 0.25), (0.14, 0.26) and
(0.19, 0.27). Dot-dashed line: isochrone with \FeH=0.14 dex and solar-scaled helium \Y=0.28. Circles: models with
(\FeH, \Y) $\equiv$ (0.14, 0.26) and \aMLT= \aMLTs $-0.20$ ($\bullet$) or \aMLTs$+0.20$ ($\circ$). 
Squares: (\FeH, \Y) $\equiv$ (0.09, 0.24), (0.14, 0.25), (0.19, 0.26) ($\square$); (\FeH, \Y) $\equiv$ 
(0.09, 0.26), (0.14, 0.27), (0.19, 0.28) ($\blacksquare$). Open triangle: (\FeH, \Y) $\equiv$ (0.14, 0.26) and OPAL EOS.
The smallness
of the error bars allows us to discriminate between rather close
values of \Y~ and to eliminate the isochrone with solar-scaled
helium.}
\label{MLvB}
\end{figure}

\begin{figure}[htb]
\resizebox{\hsize}{!}{\includegraphics*{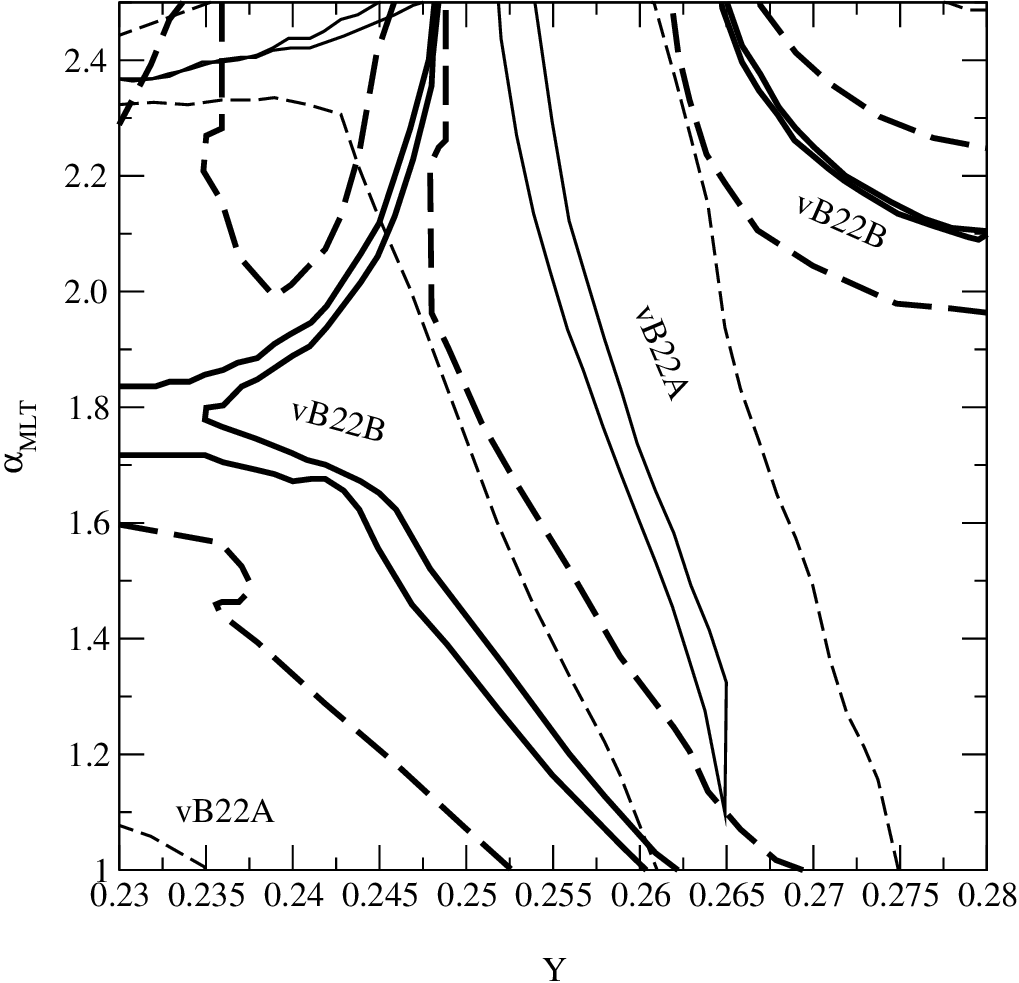}}
\caption{Contour levels in the \Y-\aMLT~plane. Continuous lines are for 
$\chi^2=(M_\mathrm{V, model}-M_\mathrm{V, obs})^2/{\sigma_{M_\mathrm{V}}}^2+(\mathrm{M_{model}-M_{obs}})^2/{\sigma_\mathrm{M}}^2 =10^{-2}$, 
dashed for $\chi^2=1$. vB22B is in bold lines.}
\label{contour}
\end{figure}
\begin{figure}[htb]
\resizebox{\hsize}{!}{\includegraphics*{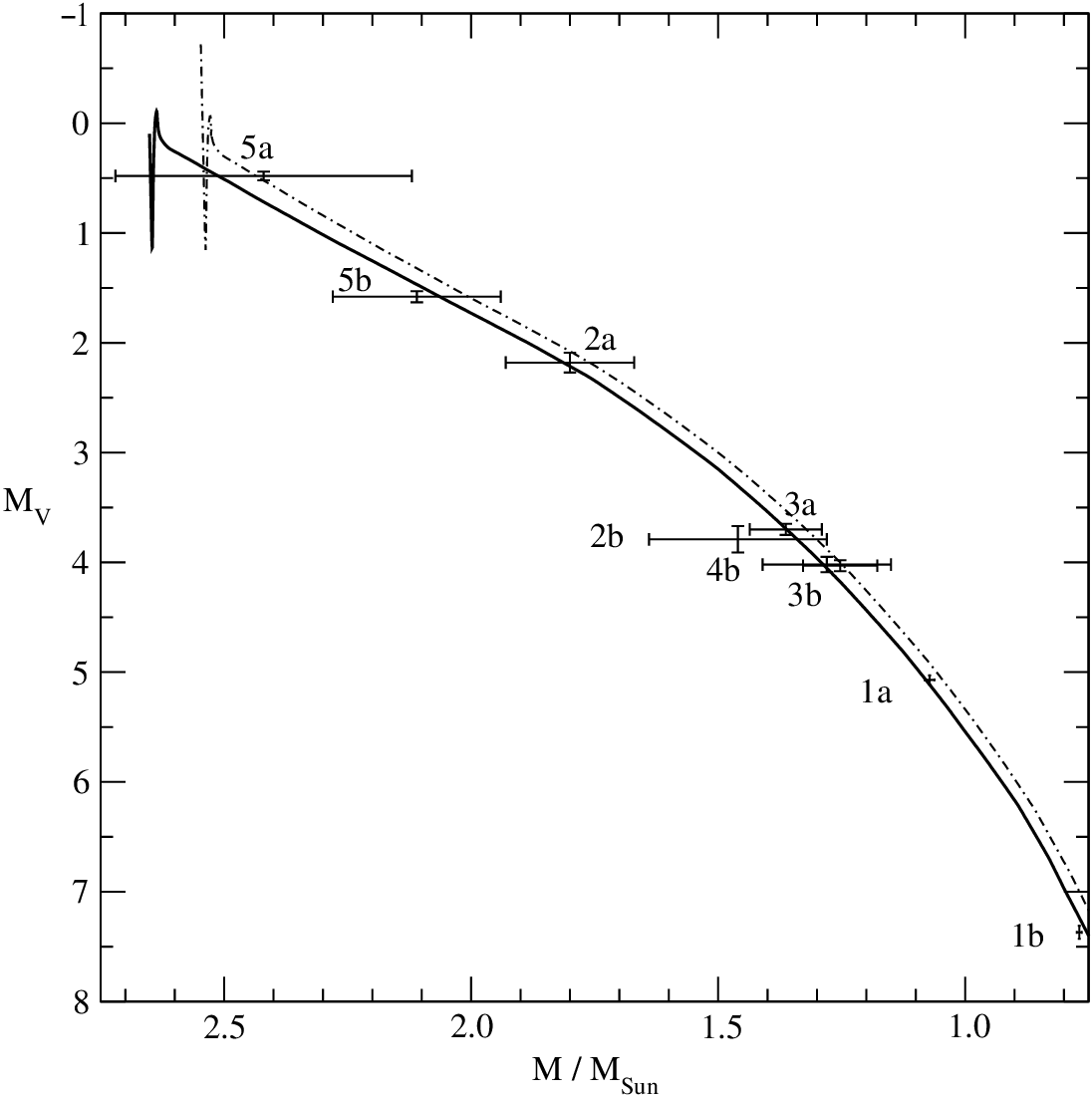}}
\caption{The Hyades M-L relation for vB22 (1), 51 Tau (2), $\phi$342 (3), $\theta^1$ Tau (4) and
$\theta^2$ Tau (5) and model isochrones aged 625 Myr. Continuous
line: three indistinguishable isochrones with 
(\FeH, \Y) $\equiv$ (0.09, 0.25), (0.14, 0.26) and
(0.19, 0.27). Dot-dash line: \FeH=0.14 dex and solar-scaled helium \Y=0.28. Here the
error bars on mass are too large to discriminate between various
helium values (vB22 excepted)} 
\label{ML}
\end{figure}

The vB22-components have the lowest and most accurate
masses ($\mathrm{\sigma_M/M} < 1 \%$) and define
the lower part of the M-L relation quite accurately.
 From the models we find that at 650 Myr (age 
of the Hyades, according to Perryman \etal \cite{per98}),
both components are only slightly evolved, the
departure from the ZAMS position being less than 0.06 mag for vB22A (0.03 mag for vB22B).  

The positions of the vB22 components in the M-L plane are
plotted in Fig.~\ref{MLvB} where we have superimposed models and 
isochrones aged 625 Myr. 
The first important conclusion that can be drawn
is that (with the present set of models) the
constraints imposed by the M-L relation are hardly compatible with a
{\emph{solar-scaled}} helium value: the (\FeH, \Y)
$\equiv$ (0.14, 0.28) isochrone lies more than $3\sigma$ above
the data. This possibility had already been suspected by Torres \etal (\cite{tor97b}). 
Even if we allow \FeH~to vary inside its error bars, we
find that an isochrone with \FeH=0.10 dex and solar-scaled helium
(not represented in the figure) sits more than $2\sigma$ above the data.
Errors on the bolometric corrections (BC) could be invoked but they should be quite high.
The BaSeL transformations applied to (\Teff, \logg)-values close to those of the vB22 system 
lead to $\mathrm{BC}_A \approx -0.1$ mag and  $\mathrm{BC}_B\approx -0.7$ mag. On the other hand, for both stars,
the BaSeL BCs do not differ by more than 0.03 mag from Alonso \etal's (\cite{alo95}) empirical corrections.
To reconcile models with solar-scaled helium with observations, the BC should be changed by 
$\Delta \mathrm{BC}_A \sim 0.15$ mag and $\Delta\mathrm{BC}_B \sim 0.40$ mag, 
which appears to be very high.
A helium content lower than the solar-scaled value is therefore favoured.

The well-known degeneracy between helium and
metallicity in the HR and M-L diagrams makes isochrones with different (\FeH,
Y) values coincide exactly. Models give $\partial M_V/\partial Y=-10\pm1$ 
and $\partial M_V/\partial\mathrm{[Fe/H]}=2.0\pm0.2$. That means that a
change in \Y~of $-0.01$ is compensated for by an increase in \FeH~by
0.05 dex. 

Several error sources bring uncertainties in the \Y~determination. 
The errors on mass and visual magnitude each lead to $\Delta Y\approx0.005$. 
An error of 0.05 mag on the bolometric correction also gives $\Delta Y\approx0.005$. 
Depending on the choice made for the Hyades \FeH-value (i.e.
spectroscopic or photometric determination) the \FeH\ contribution to the \Y-error budget will
be dominant or not: spectroscopy  gives $\Delta$\FeH=0.05 dex   and in turn
$\Delta Y=0.01$ while photometry yields $\Delta$\FeH=0.013 dex leading to $\Delta Y=0.0026$.

We also investigated how the models change when the input physics or
input parameters are modified:

\begin{itemize}
\item the component magnitudes are not changed by more than 0.001 mag if microscopic diffusion or
pre-main sequence evolution calculations are included, 
\item adopting the OPAL EOS instead of CEFF decreases 
both magnitudes by 0.02 mag,
\item changing the Eddington's grey model atmosphere into Kurucz's model only 
modifies the vB22B magnitude (by $+0.03$ mag),
\item small age variations do not change the magnitudes much 
($\partial M_V/\partial t \lesssim 10^{-4}$ mag per Myr),
\item varying \aMLT\ around \aMLTs\ yields 
$\partial M_V/\partial \alpha_\mathrm{MLT}= 0.13$ for vB22A (0.20 for vB22B),
\end{itemize}

Using the OPAL EOS, we varied \Y~and \aMLT\ to find models of vB22A and vB22B satisfying 
$\chi^2=(M_\mathrm{V, model}-M_\mathrm{V, obs})^2/{\sigma_{M_\mathrm{V}}}^2+(\mathrm{M_{model}-M_{obs}})^2/{\sigma_\mathrm{M}}^2 \leq 1$.
We took account of an error on the bolometric corrections of $\Delta \mathrm{BC}=0.05$ mag. 
Figure~\ref{contour} shows the $\chi^2$-contour levels in the \Y-\aMLT\ plane. 
If we force \aMLT\ to be in the range 1.2-2.0
 (i.e. \aMLT=\aMLTs$\pm$ 0.4, see item 4, Sect.~\ref{models}) 
and if we require that the initial helium content be the same in the two stars,
 we find solutions for \Y\ in the range 0.247-0.263. In that \Y-range we can have 
$\alpha_\mathrm{MLT,A}=\alpha_\mathrm{MLT,B}$ but the
smallest $\chi^2$ are obtained for $\Delta \alpha_\mathrm{MLT}=\alpha_\mathrm{MLT,A} - \alpha_\mathrm{MLT,B} \ne0$
with a trend for $\Delta \alpha_\mathrm{MLT}$ to increase when \Y\ decreases.

In summary, the helium content of the Hyades deduced from models based on the OPAL EOS is
$Y=0.255\pm0.009$ (\FeH\ from photometry) or $Y=0.255\pm0.013$ (\FeH\ from spectroscopy), 
that we write
$$Y= [0.255+0.2([\mathrm{Fe/H}]-0.14)]\pm0.008,$$ to highlight the effect of the \FeH-uncertainty. 
This value 
is fully compatible with \Y$=0.26\pm0.02$ derived by Perryman \etal (\cite{per98})
from the model fit of 40 ZAMS stars with spectroscopic \Teff. The changes in the input physics considered here
do not modify that result by more than $\Delta Y=0.002$.
Present models also suggest that \aMLT\ could be different in the 
two components, and higher in vB22A, but the error bars on observations are still too large to 
draw firm conclusions on that point. 

As pointed out by Lastennet \etal (\cite{las99}), models with solar-scaled helium and
a solar mixing-length ratio cannot satisfy the mass-radius (M-R) and the M-L relation simultaneously. 
From our models we find that, quite independent of the helium abundance, only vB22A models with high \aMLT\ 
(\aMLT$>$1.8) and vB22B models with very low \aMLT\ (\aMLT$<$1.0) can give the observed radii.
The simultaneous fit of the M-L and M-R relations is possible for vB22A but imposes 
hardly acceptable values of \aMLT\ in vB22B. The use of Kurucz's model atmospheres 
to fix the external boundary conditions of the models does not modify this conclusion. 

New \Teff-determinations by Cayrel de Strobel \etal (\cite{cayG01}) give
$T_\mathrm{eff,A}=5600\pm75\mathrm\ K$ and $T_\mathrm{eff,B}=4500\pm75\ \mathrm K$. 
The radii of vB22A and vB22B can be derived from the observed magnitudes and these \Teff~ 
through Stefan-Boltzmann's law and using bolometric corrections from BaSeL. We find 
$\mathrm{R_{A}=0.97\pm0.05\ R_\odot}$ and $\mathrm{R_{B}=0.64\pm0.06\ R_\odot}$. The radii based on Stefan-Boltzmann's law
are hardly compatible with and much less precise than the radii inferred from observed eclipses 
(see Table~\ref{systems}). A higher $\mathrm{R_A}$ (smaller $\mathrm{R_B}$) implies a smaller  $\alpha_\mathrm{MLT,A}$
(higher  $\alpha_\mathrm{MLT,B}$) and therefore it is easier to find models with \aMLT\ 
in the range \aMLTs$\pm$0.4 that fit both the M-L relation and radii from 
Stefan-Boltzmann's law. To draw definite conclusions on the validity of the models, 
it is important to rediscuss the observations thoroughly. In particular, 
the radii from eclipses should be firmly assessed because they are given with such high precision that they
strongly constrain the models.

Figure~\ref{ML} shows the whole M-L relation for the 5 binary
systems. Because the error bars on
the component masses are quite large (i.e. in the range 5-25
percents), no additional constraint on the helium abundance comes
from the four systems $\phi$342, 51 Tau, $\theta^1$ and
$\theta^2$ Tau. In particular, $\theta^2$ Tau A lies in the hook
region (end of the core H-burning phase) but its exact position
on the hook cannot be determined because of its large mass error bar. 

\begin{figure}[htb]
\resizebox{\hsize}{!}{\includegraphics*{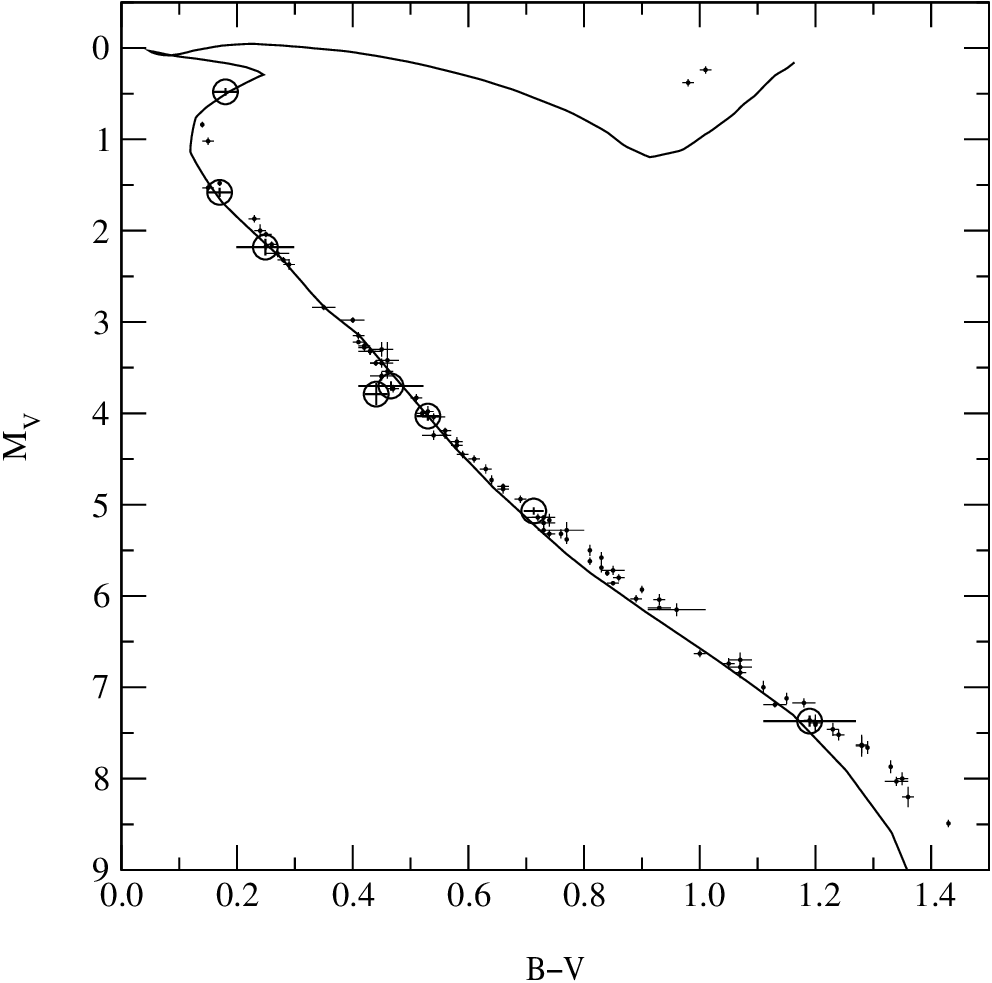}}
\caption{Hyades's HR diagram. Data are from de Bruijne \etal
(\cite{bru00}) complemented by the data of
Table~\ref{magnitudes} for four of the binary systems ($\circ$). Isochrone: 650
Myr, $Y$=0.26, \FeH=0.14, \aMLT=\aMLTs}
\label{diag_HR}
\end{figure}
\begin{figure}[htb]
\resizebox{\hsize}{!}{\includegraphics*{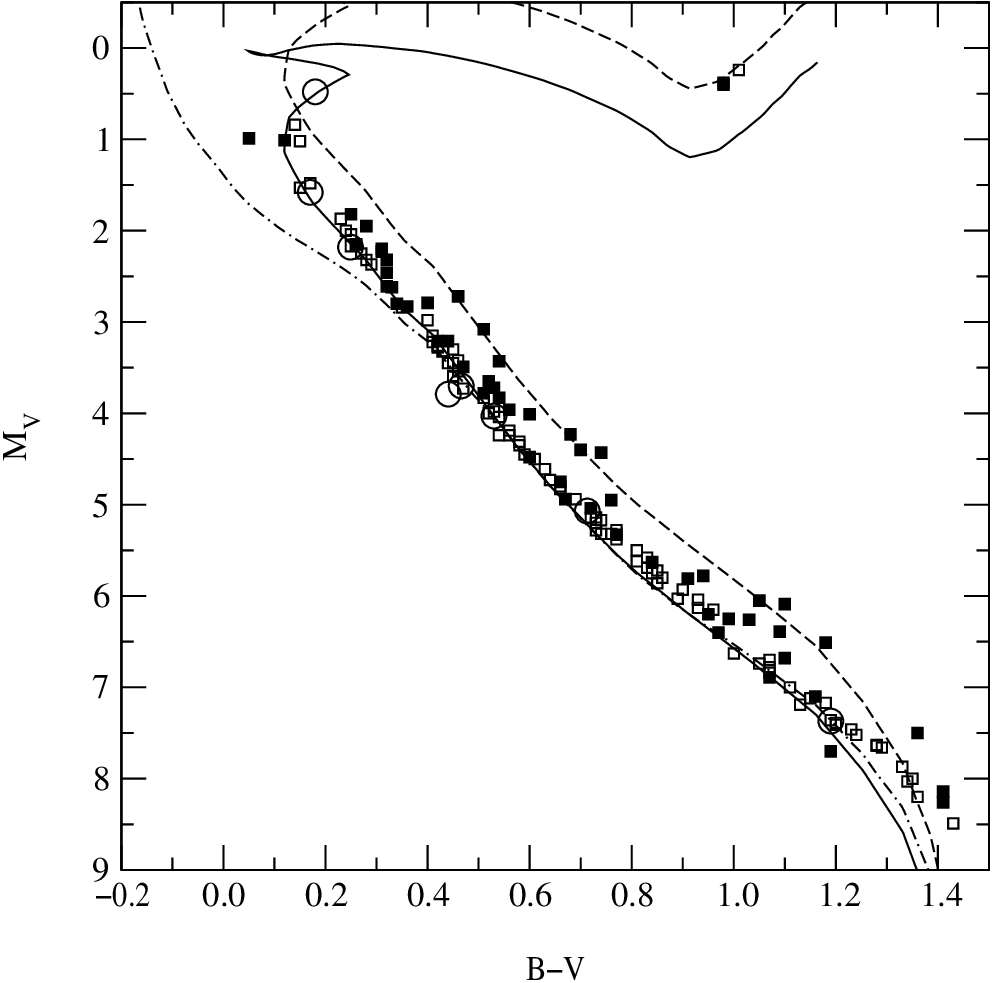}}
\caption{The HR diagram including unresolved known binaries.
Open squares are data for single stars from de Bruijne \etal
(\cite{bru00}), open circles are for the four resolved binary systems
and full squares are the unresolved binaries with dynamical
parallaxes from Dravins \etal (\cite{dra97}). Continuous line is the (650
Myr, $Y$=0.26, \FeH=0.14)-isochrone, dashed line is the same
isochrone translated by $-2.5 \log 2$ in \Mv\ giving the maximum
shift expected from unresolved binaries. The dot-dashed
isochrone is 50 Myr old.}
\label{diag_HRbin}
\end{figure}

\section{Revisited analysis of the HR diagram}
\label{HR}

Figure~\ref{diag_HR} shows the colour-magnitude diagram of 90 secure Hyades
members (de Bruijne \etal \cite{bru00}) complemented by the data of
Table~\ref{magnitudes} for four of the binary systems examined in
Sect.~\ref{M-L} above ($\theta^1$ Tau has been omitted because no
reliable information on the individual (\BV)-values is available). 
A model isochrone, aged 650 Myr, calculated with the metallicity of
the Hyades (\FeH=0.14 dex) and a helium abundance fixed by the M-L
relation (\Y=0.26) is also plotted in Fig.~\ref{diag_HR} showing that:

\begin{enumerate}
\item In the range (\BV) $\in[0.3, 1.15]$ the fit of the data is quite good, although for
(\BV) $\in[0.7, 0.9]$ some stars sit above the isochrone. 
\item For (\BV) $\gtrsim\ 1.15$ the isochrone is much too blue with respect to the
data: there is a change of slope at low mass both in the data and
in the isochrone but it occurs at lower (\BV)  in the isochrone.
\item For (\BV)$\lesssim$0.3, corresponding to the turn-off region, a few stars are
located to the right of the isochrone.
\end{enumerate}

Figure~\ref{diag_HRbin} shows the same data with, in addition, 54
known (unresolved) binaries with dynamical parallaxes from
Dravins \etal (\cite{dra97}). It shows that, except at low mass
((\BV)~$\gtrsim$ 1.15) where the slope of the isochrone is bad, 
the binaries sit well in the $\Delta M_V
\sim 0.75$ mag band, where they are expected to be.

In the following, we shall only discuss the MS and
turn-off regions. The isochrones presented here do not go far enough
to cover the giants, in the He-burning stage. In this region,
only two secure members are found, plus $\theta^1$ Tau A. For a
discussion of the constraints associated with the giant region, see
de Bruijne \etal (\cite{bru00}) and 
Castellani \etal (\cite{cas00}).

\subsection{Helium abundance}
\label{Y}

\begin{figure}[htb]
\resizebox{\hsize}{!}{\includegraphics*{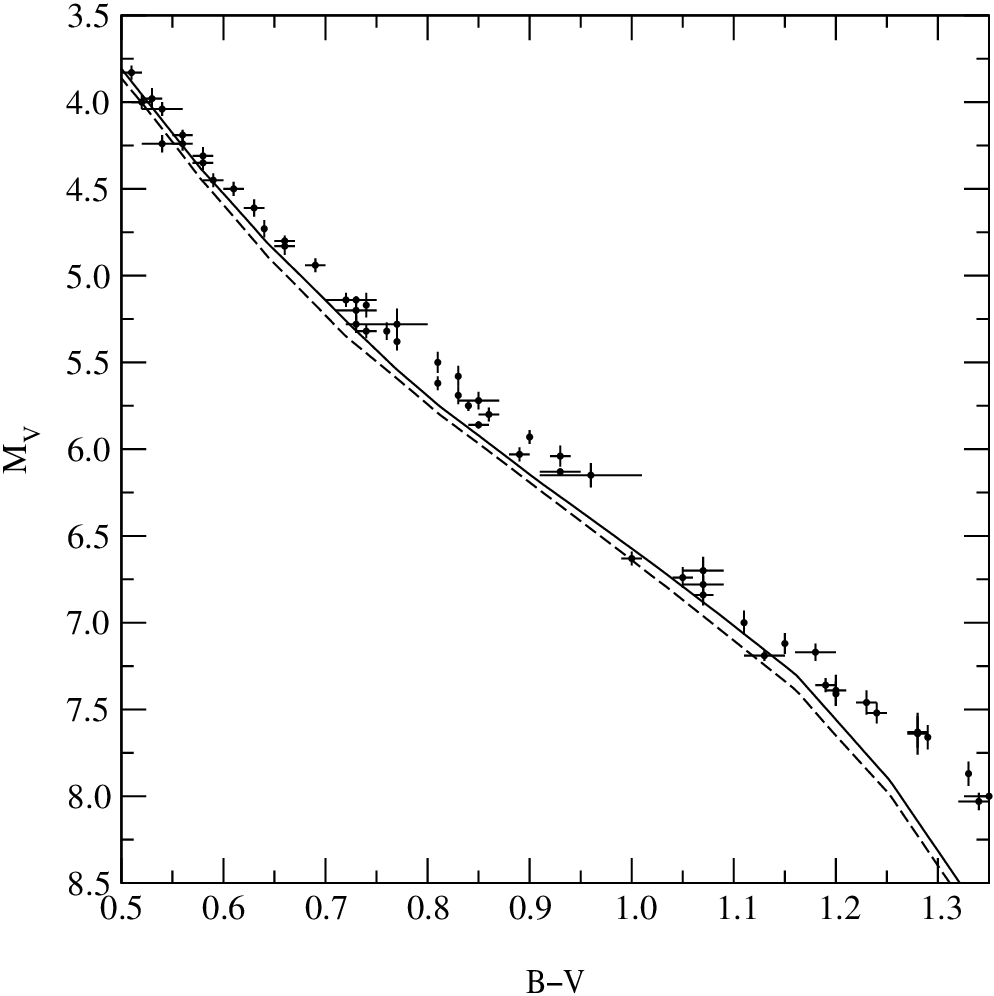}}
\caption{Helium: zoom on the non-evolved region. Isochrones aged 650 Myr, with 
\FeH=0.14 and \Y=0.26 (continuous line) or solar-scaled
helium \Y=0.28 (dash)}
\label{hel}
\end{figure}

Figure~\ref{hel} illustrates the differences between two
isochrones of same age and same metallicity (\FeH=0.14) but
different helium abundances (\Y=0.26 and solar-scaled \Y=0.28).
The isochrone that gives a better fit to
the M-L relation (i.e. \Y=0.26) also provides a better fit to
the low MS. However the constraints in the (\Mv, \BV)
plane are not so strong (the difference in \BV~ between the two
isochrones are lower than the mean error bar on \BV).

Like Castellani \etal (\cite{cas00}), we
find that the isochrone with solar scaled helium departs from the
data as soon as (B-V) $\gtrsim$ 0.6. Castellani \etal
suggest that a decrease of \aMLT\ could improve the fit 
at least for (\BV) $\in [0.7, 1.1]$. According to their Fig. 2 a
rather large decrease of \aMLT\ would be necessary (by about
$\sim$0.4). We discuss \aMLT\ changes in Sect.~\ref{slope}.

\subsection{Metallicity}
\label{Z}

\begin{figure}[htb]
\resizebox{\hsize}{!}{\includegraphics*{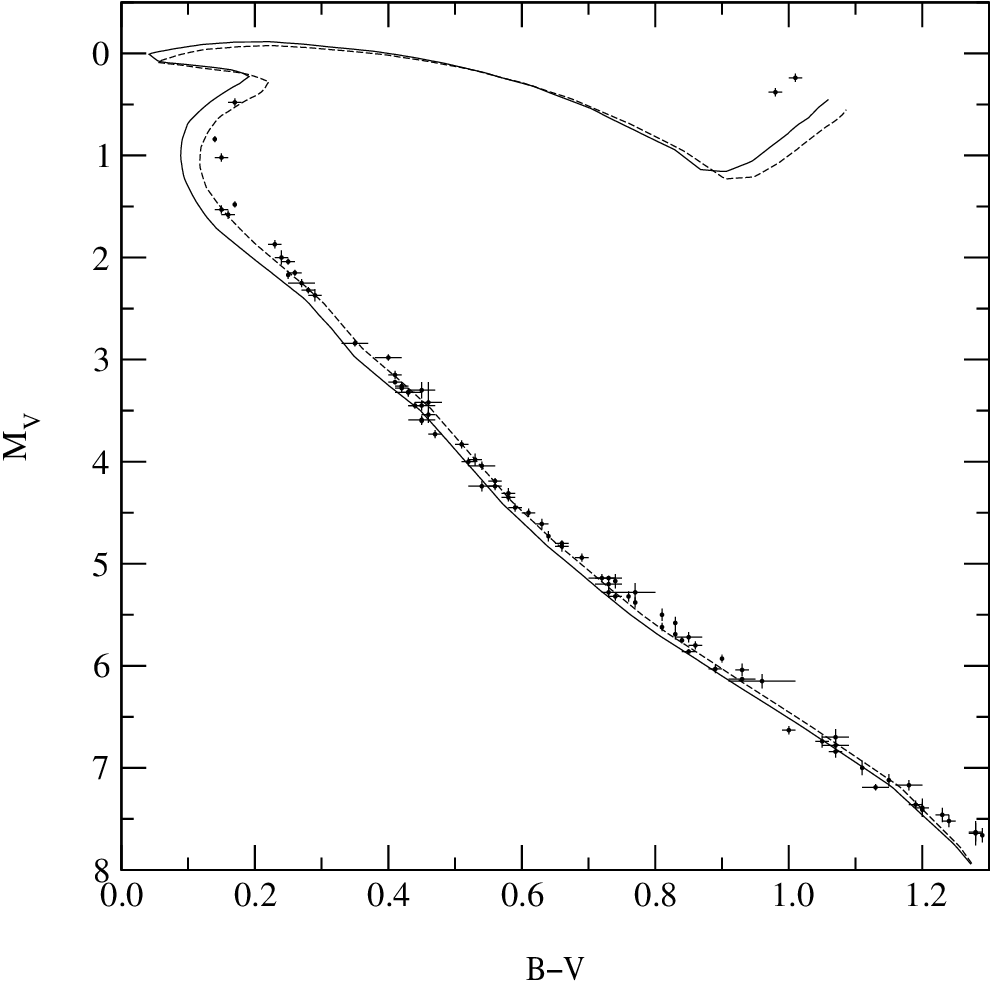}}
\caption{Two isochrones aged 650 Myr with ``extreme'' \FeH-values are plotted: 
the continuous line is 
for (\FeH=0.09, \Y=0.25), dashed for (\FeH=0.19, \Y=0.27)}
\label{metal}
\end{figure}

Fig.~\ref{metal} shows two isochrones with extreme \FeH-values (0.09 and 0.19) 
that were indistinguishable in the M-L plane (Sect.~\ref{helium}). 
In the colour magnitude diagram they differ, but by a very small
amount. On the MS, a variation of 0.05 dex in [Fe/H] changes (\BV) by less than 0.01 mag. 
The present data have $\langle \sigma_{(B-V)} \rangle \simeq 0.01$ mag which hardly allows to
discriminate between \FeH-values inside the observed range, $\Delta$\FeH=0.013 dex (0.05) from photometry (spectroscopy). 
On the other hand, an uncertainty of 0.05 dex on \FeH\ gives a 25 Myr uncertainty on the age.

\subsection{MS slope}
\label{slope}

\begin{figure}[htb]
\resizebox{\hsize}{!}{\includegraphics*{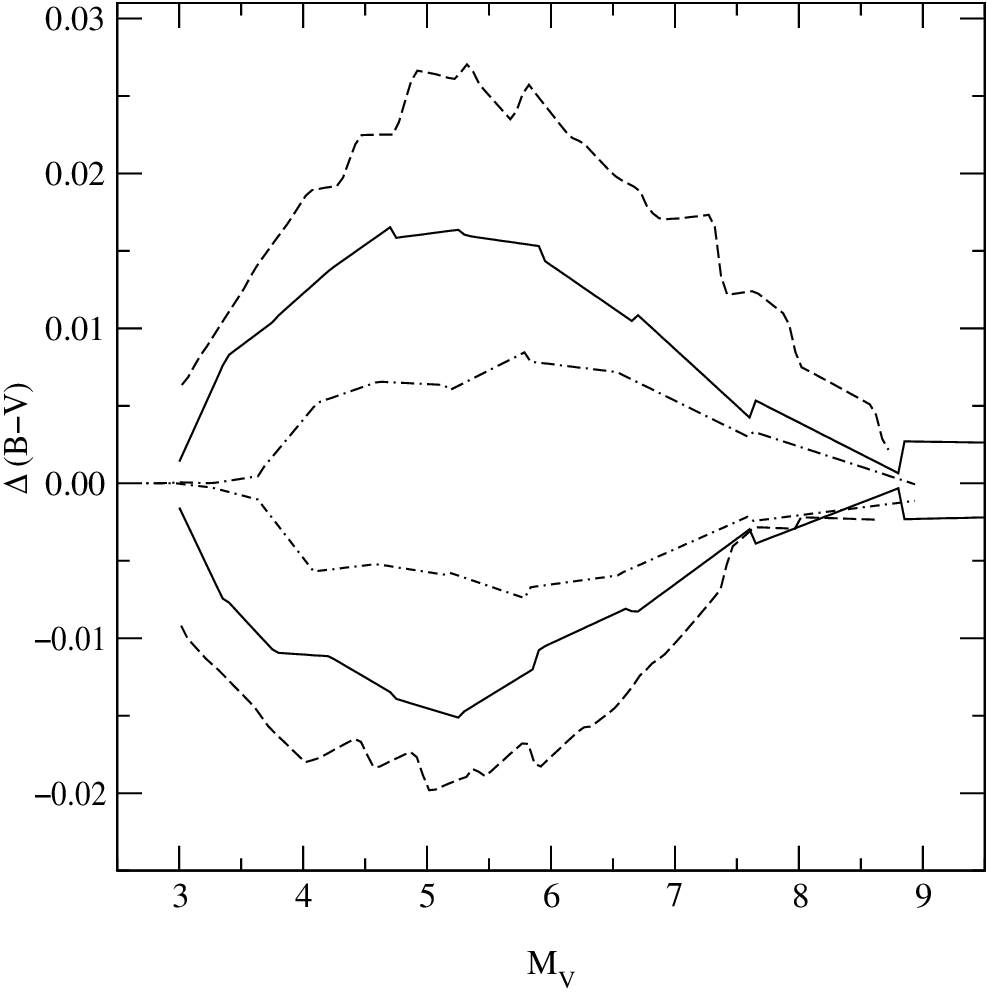}}
\caption{Effect of a change of \aMLT\ on the MS
position. $\Delta$(\BV) is the shift at constant
magnitude between an isochrone with \aMLT=\aMLTs$+\Delta$\aMLT\ and an
isochrone with \aMLTs. Positive values of $\Delta$(\BV)
correspond to $\Delta$\aMLT=~$-0.2$, negative values
to $\Delta$\aMLT=~+0.2. The continuous line is for \FeH=0.0,
dot-dash line for \FeH=~$-1.0$ and dash line for \FeH=~+0.14
dex (Hyades)}
\label{alpha}
\end{figure}

The effects of a change in \aMLT\ on the (\BV) colour index are
plotted in Fig.~\ref{alpha} as a function of \Mv~for different
values of metallicity (metal-poor \FeH=$-1.0$, solar \FeH=0.0 and Hyades
\FeH=+0.14 dex). We considered a $\pm0.20$
deviation of \aMLT\ from the solar value
(\aMLTs $\sim$ 1.6).
At any metallicity a significant shift in (\BV) is expected for
magnitudes in the range 4.-7. mag, with a maximum shift at
\Mv$\simeq$ 5.0-5.5 mag. The higher the metallicity, the higher the (\BV) shift.
 At Hyades metallicity, the (\BV) shift is in the range $\sim$0.01-0.03
mag. This effect is of the order of, and even larger than, the mean
error bar on (\BV) for the Hyades data ($\langle \sigma_{(B-V)} \rangle \simeq
0.01$ mag).

\begin{figure}[htb]
\resizebox{\hsize}{!}{\includegraphics*{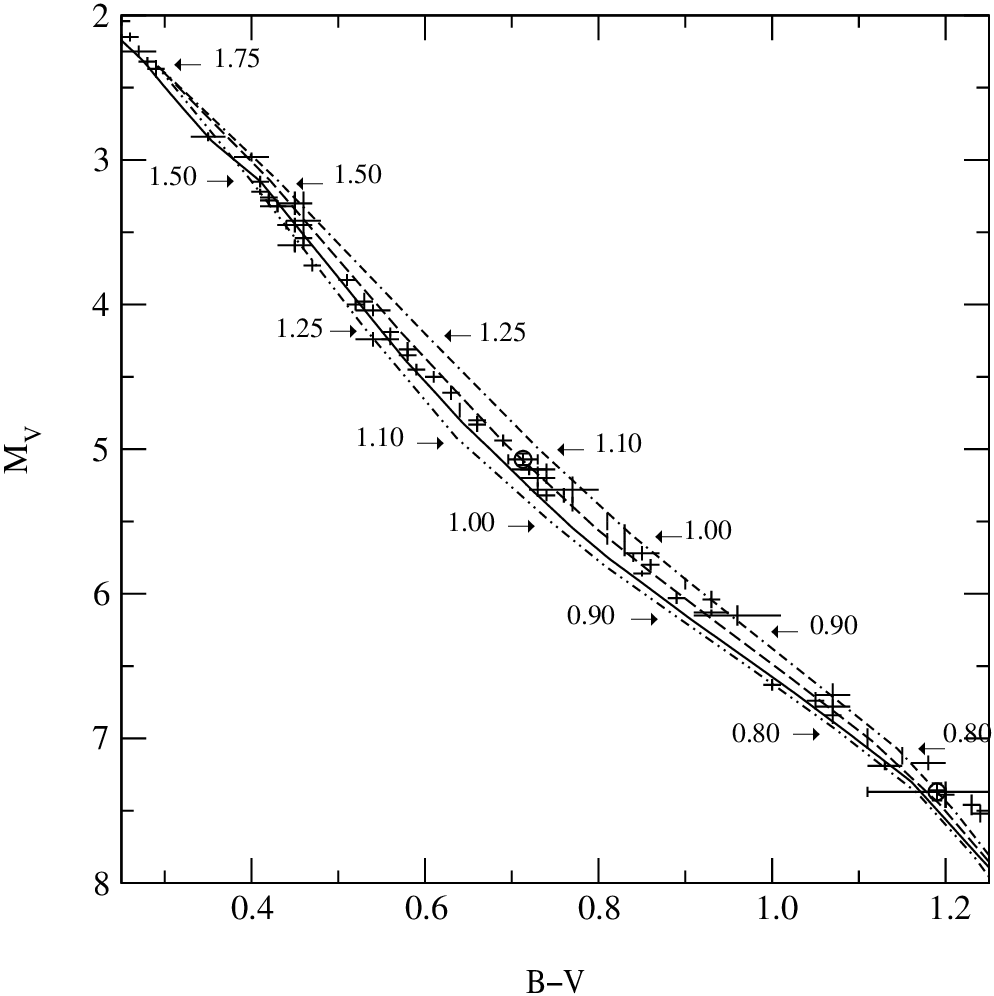}}
\caption{Effect of a change of \aMLT\ on the MS
position at Hyades metallicity. Isochrones aged 625 Myr,
\FeH=0.14, Y=0.26, are plotted: the continous isochrone is for \aMLTs, 
dashed for \aMLT=\aMLTs$-0.2$, dot-dashed for \aMLT=\aMLT$-0.4$, and
dot-dot-dashed for \aMLT=\aMLTs+0.2. The positions 
of models of different masses on the isochrones are indicated by arrows. Open circles are vB22A and vB22B}
\label{MLT}
\end{figure}

Fig.~\ref{MLT} is a zoom of the Hyades MS over the magnitude
range where the \aMLT\ changes have the largest effects.
Four isochrones with \aMLT=\aMLTs$-0.40$, \aMLTs$-0.20$, \aMLTs, 
and \aMLTs+0.20 are superimposed on the data. All stars with (\BV)$\in [0.4,1.1]$
lie in the band where \aMLTs$-0.4<$\aMLT$<$\aMLTs$+0.2$. With the present set of model
isochrones, we note a trend for stars of decreasing mass to lie on isochrones having 
decreasing \aMLT-values. This suggests \aMLT-values in the range $\sim$1.6-1.8 
for $\mathrm{M\sim1.3-1.7\ M_\odot}$, \aMLT$\approx$1.4-1.6 for $\mathrm{M\sim1.0-1.3\ M_\odot}$ and 
\aMLT$\lesssim 1.4$ below 1.0 \Msun.

   
\subsection{Lower part of the MS}
\label{lowmass}

There is a poor fit of the observed data in the cooler regions of the
HR diagram (i.e. for (\BV) $\gtrsim$1.2 mag, \Teff$\lesssim$ 4300 K).
The discrepancy in (\BV) between models and observations amounts to about
0.07 mag (0.10 mag) at (\BV)$\sim$1.3 mag (1.4 mag).

In the low-mass MS star region both the observations and the
models presented here become less secure. Errors in the equation of state
or external boundary conditions may have an important impact on the models.
On the other hand, the \BV\ color index is not the best index to be used in 
the low mass red dwarf region. 
However, we found the same kind of discrepancy in other open clusters both in (\BV) and $(V-I)$ 
(Robichon \etal \cite{rob99}, Lebreton \cite{leb00}). 
We now examine the input physics of the models.

\subsubsection{Equation of state}

\begin{figure}[htb]
\resizebox{\hsize}{!}{\includegraphics*{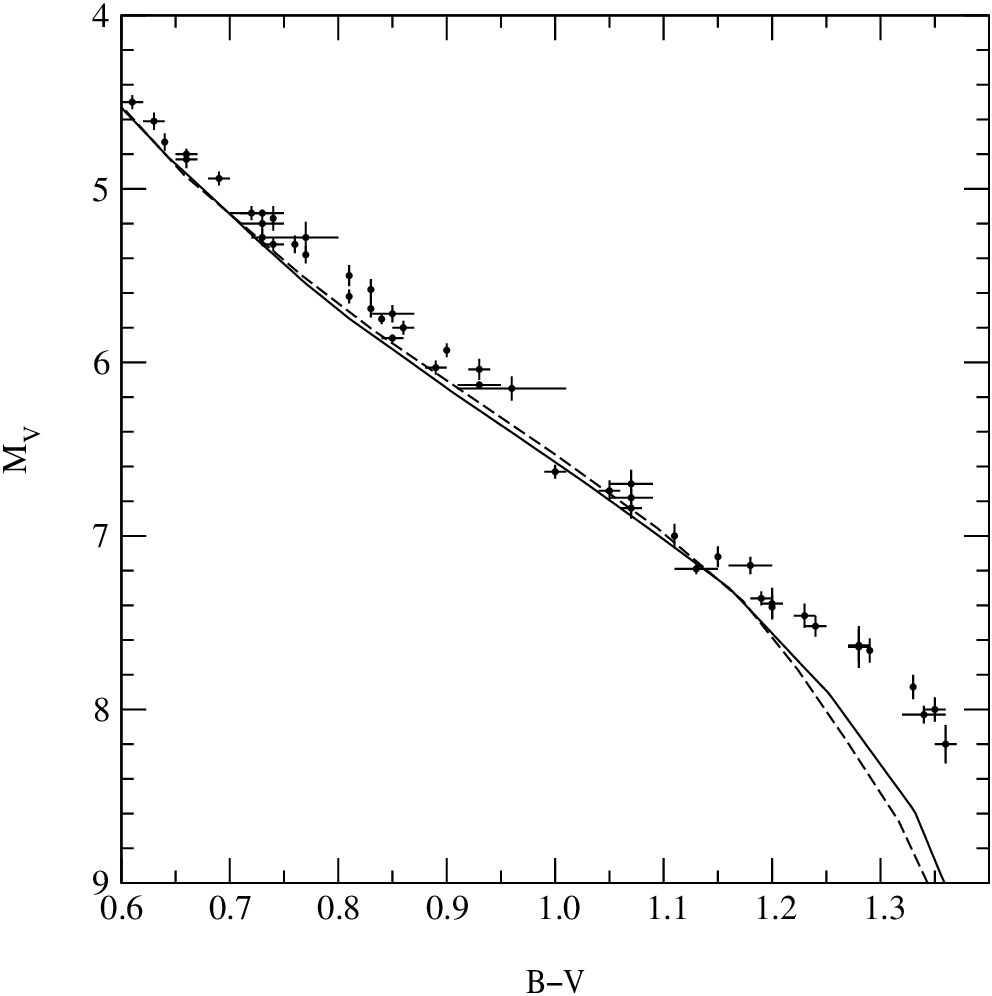}}
\caption{Effect of the EOS on the lower MS
position. The continuous line is the isochrone calculated with the CEFF EOS; dashed line with the OPAL EOS.
}
\label{eos}
\end{figure}

In most models, we used the CEFF EOS (see Sect.~\ref{models}).
Sophisticated EOSs, including collective effects, have
been designed to study the Sun (MHD EOS,
Mihalas \etal \cite{mih}; OPAL EOS, Rogers \etal \cite{rog96}) and very-low mass stars and
planets (as the SC EOS, Saumon \& Chabrier \cite{sau}).
The effects of the EOS on the models are illustrated in 
Fig.~\ref{eos} where we compare models calculated with the CEFF and OPAL EOSs.
Below 0.65 \Msun, the MS slope is steeper in models including the most sophisticated EOSs and
the fit with the observed low MS position becomes worse. 
We also calculated models with the MHD and SC EOS which exhibit the same behaviour.

Lebreton \& D\"appen (\cite{leb88}) have shown that the M-L relationship is 
very sensitive to the EOS used at low mass. We calculated models with CEFF, MHD, OPAL and SC and found that 
at solar metallicity a model of $\sim$0.50 \Msun~ is $\approx$0.75 mag brighter
when calculated with the MHD, OPAL or SC EOS instead of the CEFF EOS. 
In the Hyades, no binaries have yet been observed in this mass range,
but in the future new constraints for the physics will certainly come from low-mass binaries.  

\subsubsection{Model atmospheres}

Model atmospheres are fundamental because they fix the external
boundary conditions of the interior models and intervene in the
transformations of the model results from the (\Mbol,
\Teff)-plane to the (\Mv, \BV) colour-magnitude diagram.

The BaSeL conversions we used to convert the (\Mbol, \Teff) values to (\Mv,
\BV) are based on Kurucz's ATLAS9 model atmospheres 
down to \Teff$\sim$3500 K and on Allard \&
Hauschildt (\cite{all95}) non-grey model atmospheres beyond.
Kurucz's ATLAS9 models are known to be insufficient
below 4500 K but from preliminary tests we do not expect that conversions based on the
new model atmospheres of Hauschildt \etal (\cite{hab99}) will improve the fit
in the 4000-4500 K range.
Castellani \etal (\cite{cas00}) examined three sets of conversions 
(either empirical or based on Kurucz's models) and did not
find improvement of the slope-fitting at high (\BV). 
We also examined various empirical conversions and reach the same conclusions.

On the other hand, the boundary conditions of our models are
derived from the $T(\tau)$-laws based on Eddington's grey model atmosphere.
As demonstrated by Chabrier \& Baraffe (\cite{cha95}), because
convection penetrates into optically thin layers in the
envelope of low-mass stars, it is not correct to use grey model
atmospheres and related T($\tau$)-laws to fix the boundary 
conditions of the interior models for \Teff$\lesssim$ 4000 K. 
Chabrier \& Baraffe find that,
at solar metallicity and in the range $\sim$3000-4000 K, models 
based on the Eddington T($\tau$)-laws are hotter and brighter than 
models based on non-grey full model atmospheres, the \Teff-difference is
of $\sim$50 K at \Teff$\sim$3800 K and reaches $\sim$200 K at \Teff$\sim$ 3300 K
(their figure 2).

We have here a discrepancy in (\BV) that amounts to 0.07-0.10 mag 
at \BV$\sim$1.3 mag (1.4 mag). From the BaSeL transformations we
estimate that at \Mv$\sim$8., the effective temperature of
our models (which is $\sim$4150 K) should be reduced by more than 250 K
to increase the (\BV) index to the observed position. This appears
to be very large with respect to Chabrier \& Baraffe's results.

At high temperatures ($\gtrsim$4500K), models calculated with
outer boundary conditions taken from $T(\tau)$-laws based on 
Kurucz's model atmospheres and models using 
Eddington's grey law differ by less than 0.01 mag in (\BV).

\subsection{Turn-off region and age}
\label{TO-age}

\begin{figure}[htb]
\resizebox{\hsize}{!}{\includegraphics*{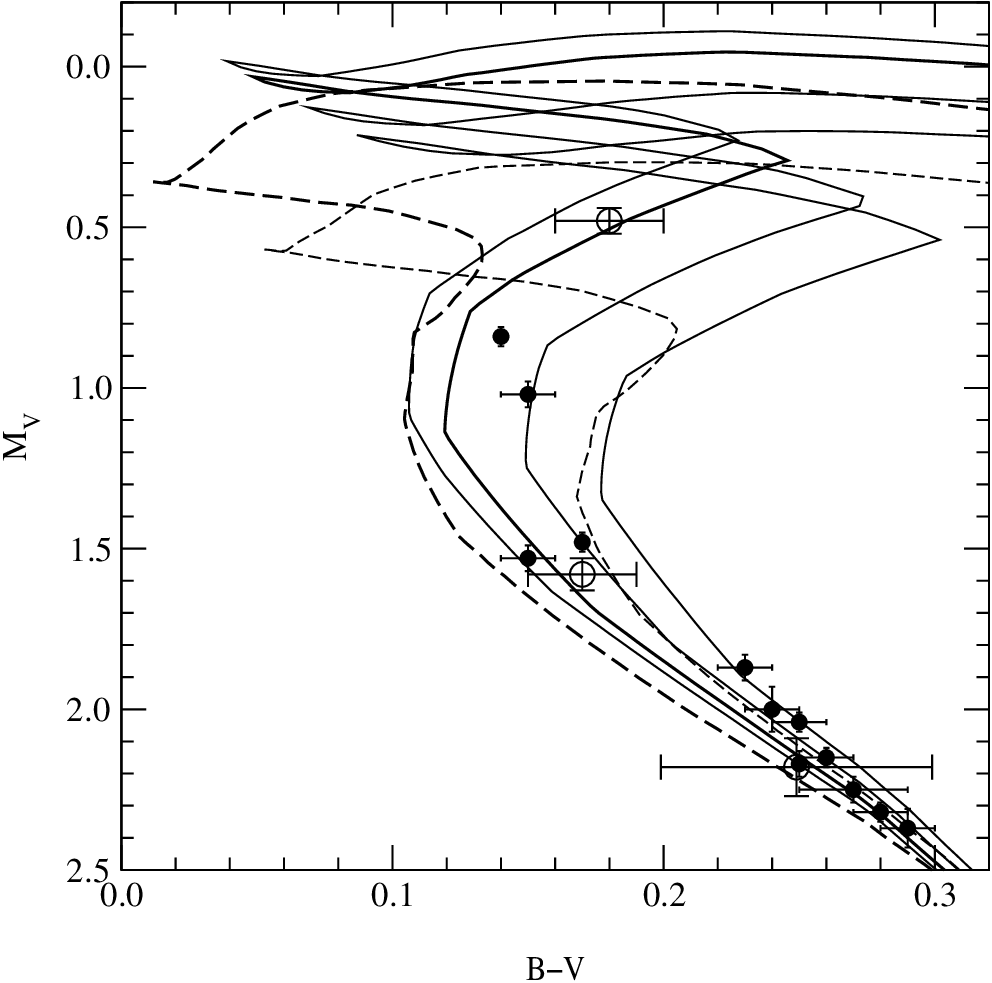}}
\caption{Zoom on the turn-off region. The observationnal 
data are the same as in Fig~\ref{diag_HR}. 
Isochrones with (\FeH=0.14, \Y=0.26) and different ages are
plotted. From left to right: continous lines are isochrones with
overshooting and aged 600 Myr, 625 Myr (bold), 700 Myr
and 750 Myr, dashed lines are isochrones without overshooting and
aged 550 Myr (bold) and 600 Myr}
\label{age}
\end{figure}
\begin{figure}[htb]
\resizebox{\hsize}{!}{\includegraphics*{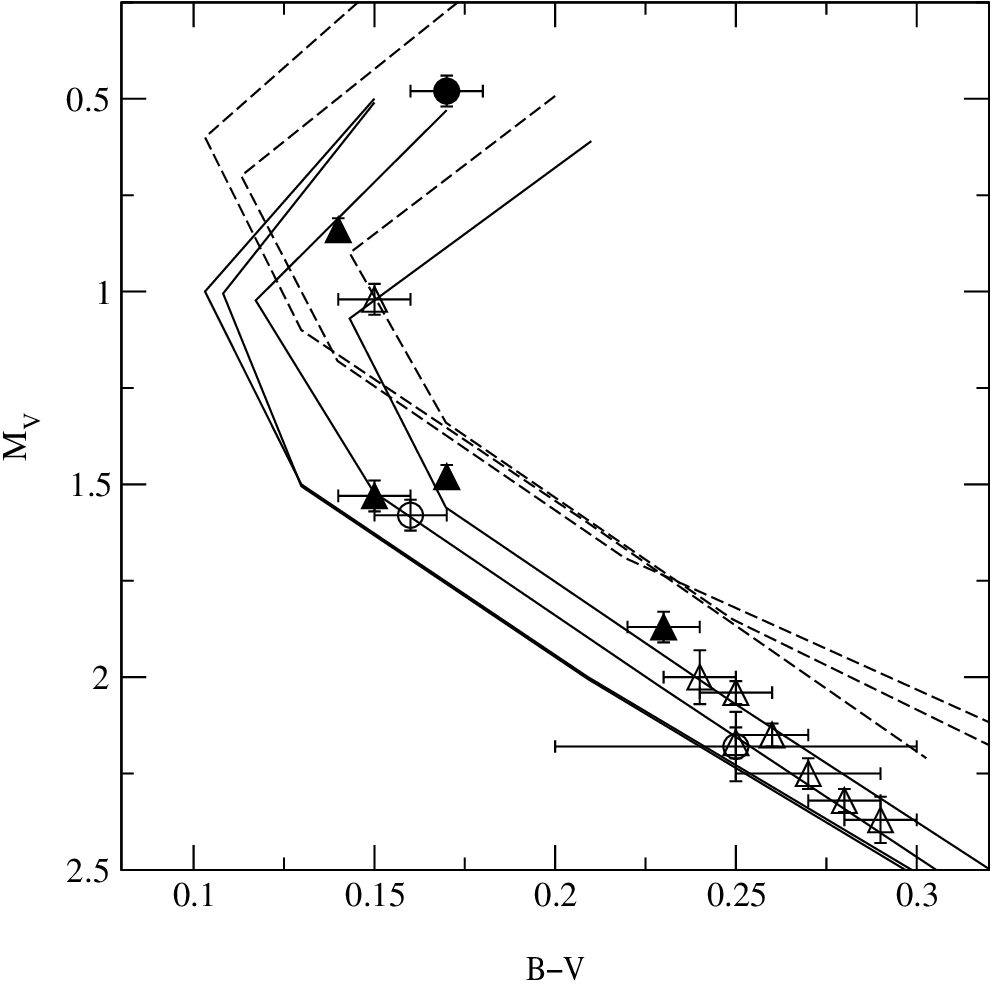}}
\caption{Effects of rotation on photometric data. Continuous lines are isochrones with from left to right
$v_e \sin i= 0, 50, 100, 150\ \mathrm{km.s^{-1}}$ and $i$=90 degrees. Dashed lines are ischrones with, from left to right (at \Mv=0.5),
$v_e \sin i= 50, 100, 150\ \mathrm{km.s^{-1}}$ and $\omega=0.95$ (see text). Open (full) symbols are Hyades with
$v_e \sin i>90\ \mathrm{km.s^{-1}}$ ($v_e \sin i<90\ \mathrm{km.s^{-1}}$). Circles are for the binaries studied in Sect.~\ref{M-L}}
\label{TO-rot}
\end{figure}

The turn-off region in the Hyades corresponds to \Mv-values below
2.5 mag. As shown in Fig.~\ref{age} there are only 12 turn-off
stars among the secure stars selected by de Bruijne \etal
(\cite{bru00}). We have added to the figure the two components of $\theta^2$ Tau
and the brighter component of the 51 Tau system.

The interpretation of observations in the turn-off region of the
Hyades are complicated by the effects of rotation and
overshooting that make either models or photometric data uncertain.

\subsubsection{Overshooting}

Hyades stars at turn-off have masses in the range 2.0-2.5 \Msun.
Therefore, they have had convective cores during their MS phase.
Overshooting of convective cores extends the mixing in central
regions: it increases the amount of hydrogen available
for H-burning and in turn, increases the MS lifetime and the MS width in the HR
diagram. The amount of overshooting is poorly known. 
Recently, Ribas \etal (\cite{rjg00}) proposed a mass dependence of the overshooting on the basis of 
the modeling of 8 eclipsing binaries. 
At $\mathrm{M\simeq2.\ M_\odot}$ and for solar metallicity, they
give $\mathrm{d_{ov}=0.17 \pm 0.05 H_p}$ which is fully compatible with $\mathrm{d_{ov}=0.2 H_p}$, the value of Schaller 
\etal (\cite{sch92}) adopted here. 

As discussed in Perryman \etal (\cite{per98}) who compared two sets
of isochrones with and without overshooting, it is possible to
fit the turn-off in both cases at different ages. This is shown
again in Fig.~\ref{age} where we have plotted isochrones of
various ages with and without overshooting. The theoretical isochrones
without overshooting are the same as in Perryman \etal
(\cite{per98}) but converted into the (\Mv, \Teff) plane by means
of the BaSeL library. For the same (\BV) at turn-off, models without overshooting
are about 100 Myr \footnote{Our models are based on $\sim$700 mass shells. 
Breger \& Pamyatnykh (\cite{bre98}) pointed out that the MS width is reduced if models are
calculated with smaller time and space steps, the effect being larger in models with
overshooting. This numerical effect could result in an overestimate of the ages in
the turn-off region of $\sim$50 Myr.} 
younger than models with $\mathrm{d_{ov}=0.2 H_p}$. Models without overshooting appear to give a poorer
fit of the position of  $\theta^2$Tau A, and at the magnitude of 
$\theta^2$Tau A, they predict a mass 2 percent larger than the mass predicted by models 
with overshooting. However, the mass of $\theta^2$ Tau A is not known
with enough precision to discriminate between models with $\mathrm{d_{ov}=0.2 H_p}$ and models without overshooting.
Furthermore, as will be discussed in Sect.~\ref{rot}, one
must also account for rotation effects in the Hyades turn-off region.

\subsubsection{Rotation}
\label{rot}

The turn-off region of the Hyades also 
corresponds to the $\delta$ Scuti instability strip. In that
region, stars are mainly rapid rotators and variable stars and,
 some of them have been identified as $\delta$-Scuti pulsators (Antonello \&
Pasinetti Fracassini \cite{ant98}). The 15 stars seen in
Fig.~\ref{age} have  $v_e \sin i$ in the range 43-175 $\mathrm{km.s^{-1}}$ 
(Bernacca \& Perinotto \cite{ber70}, 
Abt \& Morrell \cite{abt95}) and among these stars, 
10 stars have $v_e \sin i \gtrsim 90\ \mathrm{km.s^{-1}}$
($v_e$ is the equatorial velocity and $i$ the inclination of the
rotation axis with respect to the line of sight). 
In particular, 51 Tau A has measured values of $v_e \sin i$ in the
range 97-125 $\mathrm{km.s^{-1}}$, $\theta^2$ Tau A (which is a
well-known $\delta$-Scuti) has $v_e \sin i \sim 80\  \mathrm{km.s^{-1}}$
while its companion $\theta^2$ Tau B rotates even faster
with published $v_e \sin i$ in the
range 90-170 $\mathrm{km.s^{-1}}$ (Torres \etal \cite{tor97c}). 

Rotation complicates the interpretation of the HR diagram and
makes the age estimates uncertain. It affects both the models and
the photometric data.

\begin{itemize}
\item It has been known for a long time that rotation produces a
displacement of the stars in any photometric HR diagram by amounts depending on 
$v_e$ and $i$ (Maeder \& Peytremann \cite{mae72}). 
P\'erez-Hern\'andez \etal (\cite{per99}) have computed the
photometric magnitudes of rotating atmospheres. They give the
magnitude differences between non-rotating stars of given (\Teff,
\logg) and their rotating copartners, as a function of the
$i$-angle and of $\omega=\Omega/\Omega_c$ where $\Omega$ is the
star angular velocity and $\Omega_c=(8 GM/27 R_p^3)^{1/2}$ ($M$ is
the mass, $R_p$ is the polar radius). Interpolation in their Table 2 at
(\Teff, \logg)-values of the model isochrones yields the
corrections $\Delta$\Mv~ and $\Delta$(\BV)~ to apply
to our \Mv~  and (\BV)-values. We calculated the corrections for three
values of  $v_e \sin i$ (50, 100 and 150 $\mathrm{km.s^{-1}}$), and for
each $v_e \sin i$, we considered the case where $i$=90 degrees ($v_e$ and therefore $\omega$ are minimum),
and the case where $\omega=0.95$ ($v_e$ tends to the maximum allowed value, $i$ tends to its minimum possible value).

The results are plotted in Fig.~\ref{TO-rot}. When the rotation rate is close to 
maximum, the isochrone is shifted both toward higher luminosities and smaller 
effective temperatures. The shift in (\BV) is of a few tenths of a magnitude and 
does not depend much on $v_e \sin i$, 
except in the upper part of the turn-off. 
For minimum rotation rates ($i$=90 degrees), the effect of rotation is to shift the isochrones, mainly in (\BV), by an amount that increases with $v_e$. 
Rotating stars are therefore expected  to  be spread in a band
that extends on the red side of the non-rotating isochrone. 
A detailed individual analysis of each star at turn-off is beyond the scope of this work. 
However, with the help of Fig.~\ref{age}, we can estimate that
the effect of rotation on photometry could be responsible for an overestimate of the age of the Hyades by $\sim$50 Myr.
\item On the other hand, models are modified when rotation is taken into account. 
Breger \& Pamyatnykh (\cite{bre98}) showed that models with uniform rotation are 
displaced toward the right in 
the HR diagram, in the same way as models including overshooting. Furthermore, the rotation profile inside a star is related to the redistribution of 
the angular momentum by internal motions which could be generated by meridional circulation 
and shear turbulence in a rotating medium (see Zahn \cite{zah92}). Such motions might induce internal
 mixing. As shown by Talon \etal (\cite{tal97}), in the HR diagram these rotational effects also ``mimic''
 overshooting (for instance in a 9 \Msun~  star with a rotational velocity of 100 $\mathrm{km.s^{-1}}$ the effect is 
roughly equivalent to the effect of an overshooting of 0.2$\mathrm{H_p}$). 
\end{itemize}

Overshooting and rotation have similar effects on model isochrones
and therefore similar signatures in the HR diagram. These effects cannot be discriminated with the present data;
additional data on the internal structure of the considered stars are required that will hopefully
(and only) be brought by asteroseismology measurements. 
On the basis of the present data and model isochrones, the upper limit on the Hyades age is around
$\sim$ 650 Myr, the lower limit would be $\sim$500~Myr 
(no overshooting, rotation corrections on photometry included).

\section{Summary and conclusions}
\label{summary}

High accuracy observations including astrometric, photometric and spectroscopic
data have been gathered for the nearest open cluster, the Hyades. These data provide
precious information: the Hyades is the only cluster in which  
individual distances are available for a bunch of a hundred stars and in which 
individual masses have been measured in several binary systems.
These strong constraints have been used to infer the characteristics of the
cluster from the comparison with stellar models. 

The helium abundance of the cluster can be constrained by the M-L relation provided (1) the stellar data are accurate enough, 
(2) the error on the bolometric corrections needed to convert the model results in the observed M-L
plane are small and, (3) the input physics of the models is well chosen. 
The only Hyades binary system with data accurate enough to constrain the helium
abundance is vB22. We estimated the vB22 helium abundance using models including different
input physics (EOS, atmospheres, diffusion, PMS) and we allowed variations of \aMLT\  
of $\pm$0.4 around the solar value. The result, $Y=0.255\pm0.0 09$, implies
that the Hyades have less
helium than expected from their supersolar metallicity. A low helium abundance 
has already been
suggested in the past (see for instance VandenBerg \& Poll \cite{vp89}, Torres \etal \cite{tor97c}) and, 
a long time ago, on the basis of photometric measurements, Hyades stars were found to be helium
deficient with respect to field stars of same metallicity (Str\"omgren \etal \cite{sog82}). 

We pointed out that the error bar on \Y\ is much smaller if \FeH\ is taken from photometry
(internal error bar on \FeH\ around 10 percent, Grenon \cite{gre00}) rather than from spectroscopy 
(internal error bar on \FeH\ around 35 percent, Cayrel \etal \cite{cayG97}).
Progress will come from a better description of the so-called metallicity by the future determinations of
the individual abundances of the main metallicity contributors. The C, N, O abundances are presently 
poorly known while the Ne abundance is unknown (R. Cayrel, private communication).
On the other hand, it would be worthwhile to improve the mass values in the 
four other binary systems 
which are presently not good enough to constrain models (and helium) significantly.

An upper limit of $\sim$650 Myr on the cluster age can be estimated from the positions of stars 
at turn-off. It is difficult to estimate precisely the age of the Hyades because at turn-off
rapidly rotating stars are found which had a convective core on the MS. 
Rotation affects the photometry and makes the observed HR diagram positions uncertain, 
whereas internal rotation and overshooting are responsible for internal 
mixing that may change the model course in
the HR diagram. As a result, the age of the Hyades might be in the range 500-650 Myr. An improvement of
the mass of $\theta^2$Tau A would certainly better constrain the overshooting
by anchoring the star more precisely on its isochrone. 
Asteroseismological measurements and analysis are required to get a
better understanding of the transport processes inside stars (Goupil \etal \cite{gou00}).

The observed MS slope, in the region of the HR diagram corresponding to masses in the range 
0.8-1.5 \Msun, is quite well-defined and provides first tests of the \aMLT-value to be 
used in models. It suggests that \aMLT\ could be a slowly decreasing function of mass: \aMLT\ 
would decrease from values of $\sim$1.8-1.9 at high mass to values of $\sim$1.2-1.4 at 
low mass.
The confirmation of that trend would require extremely precise radii or \Teff\  
for individual stars along the MS. Presently, both radii and \Teff\ are 
only available for vB22 (only vB22A sits in the HR diagram region sensitive to 
\aMLT) and since they do not agree,
we have not been able to draw firm constraints on \aMLT\ from these data.

In the very low mass region of the HR diagram, models are definitely too blue. This region 
provides tests on the model external boundary conditions and on the equation of state and 
the knowledge of a few masses would help to understand the origin of the discrepancy.

The Hyades stars are now better understood because they are  fully seen as individuals and because of the
considerable recent progress in the description of the physics of the stellar plasma (opacities, EOS).
In  1989, VandenBerg \& Poll (\cite{vp89}) used stellar models 
to infer the Hyades distance modulus by means of the MS fitting technique  
assuming that the helium content was solar. Now that we have removed the distance problem,
 we are able to test finer details in the models. We can hope that in the decade 
to come, a substantial number of
open clusters will come to be known at the level of accuracy presently reached for the Hyades and even
better. Hipparcos provided mean distances to a few open clusters which already 
give rise to many questions. However, 
the discussion of the helium abundance in the nearer clusters is limited because 
errors sources can be important (color-calibrations and bolometric corrections,
distances, metallicity) and because of the lack of well observed binaries 
(Robichon \etal \cite{rob99}, Lebreton \cite{leb00}). 
In the future, GAIA will determine individual distances of stars in many clusters
together with information on their abundances (Perryman \etal \cite{per97}) : it is expected that about
120 clusters, located to within 1 kpc, will reach to the level of accuracy or even better than what is reached today
in the Hyades.
GAIA will also discover binary systems in which it will measure masses accurately (at least ten binary systems are
expected per cluster). Other observational devices will also provide accurate masses and radii in the near future. 
Therefore, we can expect that we shall go on constraining the models to 
progress in the understanding of the chemical and dynamical evolution of our Galaxy.

\begin{acknowledgements}
We warmly thank D. Dravins, L. Lindegren and J. de Bruijne 
who provided their parallax data. We are grateful to 
A. Baglin, G. Cayrel de Strobel, R. Cayrel,
H.-G. Ludwig, F. Arenou, M.J. Goupil and E. Michel for many fruitful
discussions. We thank the referee Dr. A.A. Pamyatnykh for interesting
remarks and information.
Y. Lebreton acknowledges University of Rennes 1 for
working facilities.	This work was partially supported by the ``Convénio ICCTI Embaixada de
França, number 060/B0'' and by the project ``ESO/P/PRO/12128/1999'' from ``Fundação
para a Ciência e Tecnologia''.
\end{acknowledgements}

\end{document}